\documentclass[aps,prb,reprint,longbibliography,superscriptaddress]{revtex4-2}
\usepackage{mathrsfs}
\usepackage{amsmath,gensymb}
\usepackage{amsfonts}
\usepackage{amssymb}
\usepackage{amsthm}
\usepackage{graphicx}
\usepackage{natbib}
\usepackage{color}
\usepackage{hyperref}
\usepackage{bm}
\usepackage[caption=false]{subfig}
\usepackage{verbatim}
\usepackage[normalem]{ulem}

\begin{document}

\title{Ultrastrong magnon-photon coupling in
superconductor/antiferromagnet/superconductor 
heterostructures at terahertz frequencies}

\author{V. M. Gordeeva}
\thanks{These authors contributed equally}
\affiliation{Moscow Institute of Physics and Technology, Dolgoprudny, 141700 Moscow region, Russia}

\author{Yanmeng Lei}
\thanks{These authors contributed equally}
\affiliation{School of Physics, Huazhong University of Science and Technology, Wuhan 430074, China}

\author{Xiyin Ye}

\affiliation{School of Physics, Huazhong University of Science and Technology, Wuhan 430074, China}

\author{G. A. Bobkov}
\affiliation{Moscow Institute of Physics and Technology, Dolgoprudny, 141700 Moscow region, Russia}

\author{A. M. Bobkov}
\affiliation{Moscow Institute of Physics and Technology, Dolgoprudny, 141700 Moscow region, Russia}

\author{Tao Yu}
\email{taoyuphy@hust.edu.cn}
\affiliation{School of Physics, Huazhong University of Science and Technology, Wuhan 430074, China}

\author{I. V. Bobkova}
\email{ivbobkova@mail.ru}
\affiliation{Moscow Institute of Physics and Technology, Dolgoprudny, 141700 Moscow region, Russia}
\affiliation{National Research University Higher School of Economics, 101000 Moscow, Russia}

\begin{abstract}
We predict the realization of ultrastrong coupling between magnons of antiferromagnets and photons in superconductor/antiferromagnet/superconductor heterostructures at terahertz frequencies, from both quantum and classical perspectives. The hybridization of the two magnon modes with photons strongly depends on the applied magnetic field: at zero magnetic field, only a single antiferromagnetic mode with a lower frequency couples to the photon, forming a magnon-polariton, while using a magnetic field activates coupling for both antiferromagnetic modes. The coupling between magnon and photon is ultrastrong with the coupling constant $\sim$ 100 GHz exceeding 10\% of the antiferromagnetic resonant frequency. The superconductor modulates the spin of the resulting magnon-polaritons and the group velocity, achieving values amounting to several tenths of the speed of light, which promises strong tunability of magnon transport in antiferromagnets by superconductors.    
\end{abstract}

\maketitle

\section{Introduction}
\label{intro}

Quantum magnonics is an emerging field of research focusing on the coherent coupling of magnons and photons. It promises to enable hybrid quantum platforms \cite{Tabuchi2015,Lachance-Quirion2020}, magnon-based memory systems \cite{Zhang2015}, and microwave-to-optical quantum transducers \cite{Hisatomi2016}. 
A primary challenge hindering advancements in quantum magnonics is the inherently weak coupling strength between individual spins and photons. To overcome this limitation in microwave cavities,  researchers exploit the Dicke cooperative coupling relation,
$g = g_s \sqrt{N}$,
where \( g_s \) is the single-spin coupling strength and \( N \) is the number of spins in the system \cite{Lachance-Quirion2019,Huebl2013,Tabuchi2014,Zhang2014,Kirton2019}. This approach has led to the demonstration of the strong \cite{Huebl2013,Tabuchi2014,Zhang2014} and, more recently, the ultra-strong \cite{Flower2019,Rameshti2015,Bourhill2016} photon-magnon coupling regimes, although in truly macroscopic systems with dimensions of several millimeters. Specifically, the ultra-strong coupling and deep-strong coupling regimes are defined by the conditions that the coupling ratio $g/\omega>0.1$ and $>1$, respectively. Here $\omega$ is the characteristic bare frequency of the coupled excitations.
The ultra-strong coupling regime dramatically alters the energy levels and dynamics of a coupled system \cite{Kockum2019,Qin2024}. This makes it particularly valuable for probing fundamental quantum effects and developing novel quantum technologies. 

For practical applications, on-chip integration of such quantum systems is essential, which requires a high single-spin coupling strength \( g_s \). Several successful on-chip implementations of both the strong \cite{Li2019,Hou2019} and ultra-strong \cite{Golovchanskiy2021,Golovchanskiy2021_2} photon-to-magnon coupling regimes were reported. In Refs.~\cite{Golovchanskiy2021,Golovchanskiy2021_2} the ultra-strong and approaching deep-strong photon-to-magnon coupling was demonstrated in superconductor/ferromagnet/superconductor heterostructures. The coupling ratio $g/\omega \approx 0.6$ with the single-spin coupling strength $g_s \approx 350$ Hz and cooperativity about $10^4$ was reported. Later, these data received a theoretical explanation \cite{Silaev2022,Qiu2024}. The mechanism of such ultrastrong coupling is based on the fact that magnon excitations of the ferromagnetic layers are accompanied by Meissner supercurrents in the superconducting layers, which strongly modify and enhance the microwave magnetic field inside the superconducting resonator.

On the other hand, antiferromagnets have recently emerged as attractive candidates to replace ferromagnets in spintronic devices because they are robust against magnetic perturbations, generate weaker stray fields, and exhibit ultrafast dynamics in the terahertz frequency range \cite{Baltz2018,Jungwirth2016,Brataas2020}. For this reason achieving strong and ultra-strong magnon-photon coupling regimes is an important task of antiferromagnetic quantum magnonics. Strong coupling between magnons in antiferromagnets and photons is a challenge in cavity magnonics since the magnons typically have a frequency of terahertz, while the cavity supporting terahertz photons is rare. Ultrastrong coupling is more difficult. Nevertheless, the strong-coupling regime, characterized by a bright and a dark magnon-polariton, was predicted in antiferromagnetic microcavities by Yuan et al. \cite{Yuan2017}. Further, the magnon-polaritons in THz cavities were observed experimentally \cite{Kritzell2023,Grishunin2018,Sivarajah2019,Chen2025,Bialek2020,Bialek2021,Metzger2022,Baydin2023,Blank2023}  and the coupling strength of the order of $0.015$~THz corresponding to the strong-coupling regime has been reported \cite{Kritzell2023,Grishunin2018}.

In this work, we propose a realization of ultrastrong coupling between magnons of antiferromagnets and photons in superconductor/antiferromagnet/superconductor heterostructures at terahertz frequencies. Our findings reveal that in the absence of an external magnetic field, the coupling is selective, involving only one magnon mode to form a magnon-polariton. Under an applied magnetic field, however, both antiferromagnetic modes become coupled to the photon. The coupling is comprehensively analyzed from both quantum and classical perspectives, with particular focus on the spin and group velocity of the resulting magnon-polaritons, showing the strong modulation of magnon transport in antiferromagnets by superconductors.

\section{Model}
\label{sec:model}

\begin{figure} [h]
    \centering
    \includegraphics[width=0.85\linewidth]{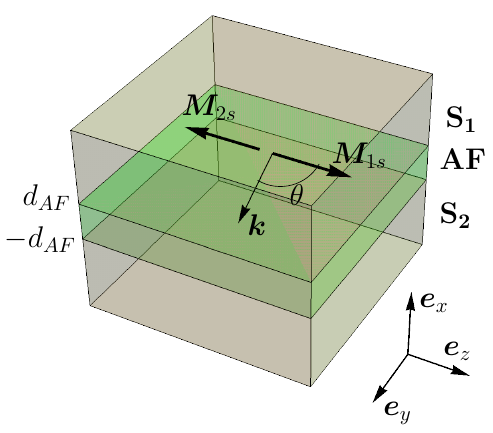}
    \caption{Sketch of the S/AF/S system. The equilibrium magnetizations of the two sublattices $\bm M_{1,2s}$ are oriented along the $z$ axis. The magnon with wavevector $\bm k$ can propagate at an angle $\theta$ with respect to $\bm M_{1s}$.}
    \label{fig:setup}
\end{figure}

As depicted in Fig.~\ref{fig:setup}, we consider a superconductor/antiferromagnet/superconductor (S/AF/S) heterostructure. 
An insulating easy-axis antiferromagnet with thickness $2d_{AF}$ is sandwiched between two thick superconductors with thicknesses exceeding the effective penetration depth $\lambda_{\rm{eff}}$ for the magnetic field at THz frequencies, which is a typical range of the antiferromagnetic magnon frequencies. Detailed discussion of $\lambda_{\rm{eff}}$ is provided below. The system hosts both photonic modes
and magnons, which are coupled via the electromagnetic interaction, producing magnon-polaritons. We assume $\lambda_{\rm{eff}}\sim(1\div3)~\mu$m and $d_{AF}\lesssim \lambda_{\rm{eff}}$.  Since the typical wave numbers of terahertz photons are $k \lesssim 0.1$ mm$^{-1}$, in the case under consideration, we can use the approximation $k \lambda_{\rm{eff}} \ll 1$. The coordinate system is chosen so that the $x$ axis is perpendicular to the S/AF interfaces with $x=0$ in the middle of the AF layer. 

The net magnetization of the antiferromagnet is $\bm M=\bm M_1+\bm M_2$, where $\bm M_{1,2}$ are the magnetizations of the two sublattices, denoted by indices "1" and "2". The equilibrium magnetizations of the two sublattices $\bm M_{1,2s}$ are oriented along the in-plane $z$ axis: $\bm M_{1s}=-\bm M_{2s}=M_s\bm e_z $. In the linear response regime magnons can be described by the transverse fluctuations of the magnetizations of the two sublattices as $\bm M_{1(2)\omega}(\bm \rho, t)=\tilde {\bm M}_{1(2)} e^{i\bm k\cdot\bm \rho-i\omega t}$ with amplitudes $\tilde {\bm M}_{1(2)}=\tilde M_{1(2)x}\bm e_x+\tilde M_{1(2)y}\bm e_y$, wavevector $\bm k=k_y \bm e_y+k_z \bm e_z$, in-plane radius vector $\bm \rho=y \bm e_y+z \bm e_z$, and frequency $\omega$. Thus, the full magnetizations of the two sublattices take the form $\bm M_{1(2)}=\bm M_{1(2)s}+\bm M_{1(2)\omega}$, and the net magnetization is $\bm M=\tilde {\bm M}e^{i\bm k\cdot\bm \rho-i\omega t}$, $\tilde {\bm M}=\tilde {\bm M}_{1}+\tilde {\bm M}_{2}$. We introduce the angle $\theta$ between the wave vector $\bm k$ and the equilibrium magnetization $\bm M_{1s}$, i.e., $k_z=k\cos\theta$ and $k_y=k\sin\theta$ with $k=|\bm k|$.

\section{qualitative picture of magnon-polaritons}
\label{sec:qualitative}

Before delving into detailed quantum and classical derivations, we present here a concise qualitative description of magnon-polariton physics in S/AF/S heterostructures. This overview highlights the key novel features of our proposal: the achievement of ultrastrong coupling at terahertz frequencies with exceptionally high cooperativity, the magnetic-field-tunable selectivity of the coupling, and the emergence of hybrid quasiparticles with remarkable properties, including a non-integer and \(k\)-dependent spin with high group velocity.

\subsection{Formation mechanism} 

The core effect originates from the electromagnetic interaction between terahertz magnons in the antiferromagnetic (AF) layer and the photonic Swihart mode of the superconducting (S) resonator. The formation of magnon-polaritons requires three essential ingredients: the Swihart photon, the bare magnon, and their mutual dipolar interaction.

\subsubsection{Swihart photon mode}

The eigenfrequency of the Swihart mode in the superconducting resonator is given by \cite{Swihart1961}
\begin{align}
\Omega_s({\bm k})=\sqrt{\frac{d_{AF}}{\mu_0\varepsilon_{AF}(d_{AF}+\lambda_{\mathrm{eff}})}}|{\bm k}|,
\label{qu:Swihart_mode}
\end{align}
which vanishes in the limit \(\lambda_{\mathrm{eff}} \to \infty\). Here $\mu_0$ is the vacuum magnetic permeability and $\varepsilon_{AF}$ is the dielectric constant of the antiferromagnet. Strictly speaking, a resonant mode persists even in a normal-metal environment, but the physics of magnon-polaritons is then strongly modified by dissipation \cite{Qin2024}. For wavevectors aligned along the \(z\)-axis (\(\bm k = k \bm e_z\)), the magnetic field of the Swihart mode is polarized along the \(y\)-axis, as illustrated in Fig.~\ref{fig:qualitative}(a). Further details on the Swihart mode and its quantization are provided in Sec.~\ref{sub:swihart}.

\begin{widetext}
\begin{center}
\begin{figure}[h]
    \centering
    \includegraphics[width=0.87\linewidth]{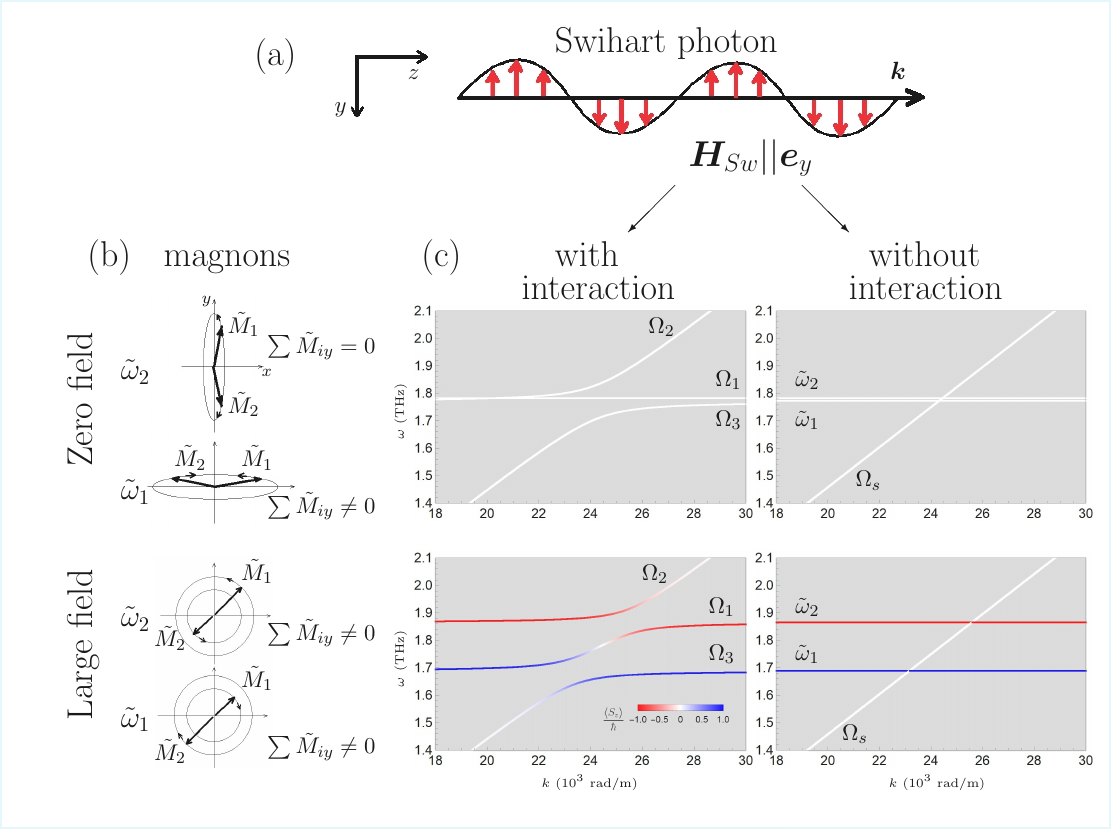}
    \caption{Magnon-polaritons in S/AF/S heterostructures. (a) Magnetic field profile of the Swihart photon propagating in the superconducting resonator. (b) Magnetization precession of the antiferromagnetic magnon eigenmodes at zero and at a large applied magnetic field $\bm H_0 = H_0 \bm e_z$. (c) Dispersion of the hybridized eigenexcitations in the S/AF/S heterostructure: noninteracting case (photons and magnons decoupled, right panel) and interacting case (magnon-polariton formation, left panel). The averaged projection of the eigenexcitation spin $\langle S_z \rangle$ onto the $z$-axis is shown by color.}
    \label{fig:qualitative}
\end{figure}
\end{center}
\end{widetext}

\subsubsection{Magnons}

The eigenmodes of bare magnons are derived in Sec.~\ref{sub:magnons}. Neglecting demagnetization effects, the two eigenfrequencies correspond to the well-known antiferromagnetic resonance modes for an easy-axis antiferromagnet \cite{Rezende2019}:
\begin{align}
\hbar \omega_{1,2}^b=\mu_0\gamma\hbar \left(\sqrt{(K+J_{AF})^2-J^2_{AF}}\pm H_0\right),
\label{qu:eigen_bare}
\end{align}
where \(-\gamma\) is the electron gyromagnetic ratio, \(J_{AF}\) is the antiferromagnetic exchange constant, \(K\) is the easy-axis anisotropy constant, and \(H_0\) is the external magnetic field along the N\'eel vector. The two modes are degenerate when $H_0=0$ due to the $PT$ symmetry.

In the S/AF/S geometry, the dynamic magnetization of the AF layer generates stray magnetic fields, which break the $PT$ symmetry and $U(1)$ symmetry around the N\'eel vector. It in turn induces Meissner screening currents in the adjacent superconducting layers. This current-field configuration strongly modifies the effective dipolar interaction within the AF medium. The demagnetization field \(\bm H(\bm r,t)=\tilde{\bm H}(x)e^{i\bm k\cdot\bm \rho-i\omega t}\) governs this interaction via the term
\begin{align}
\hat{H}_{\mathrm{dem}}=-\mu_0\sum_{i=1,2}\int d{\bm r} \hat{\bm M}_i({\bm r})\cdot {\bm H}(\bm r,t).
\end{align}
The demagnetization field, accounting for the influence of the superconducting environment, is calculated in Sec.~\ref{sec:dipolar} and can be expressed in terms of a demagnetization tensor:
\begin{align}
\hat H=-\hat N \hat {M},\quad
\hat H=\begin{pmatrix}
\tilde H_{x}  \\
\tilde H_{y}
\end{pmatrix},\quad
\hat {M}=\begin{pmatrix}
\tilde M_{x}  \\
\tilde M_{y}
\end{pmatrix},
\label{qu:demagnetization}
\end{align}
with
\begin{align}
\hat N=\begin{pmatrix}
1  & 0 \\
0 & N_y (k)
\end{pmatrix},
\label{qu:N_tensor}
\end{align}
where \(N_y(k=0) \equiv N_{y0} = d_{AF}/(d_{AF}+\lambda_{\mathrm{eff}})\) and assumes a more complex form for \(k \neq 0\). The bare magnons correspond to \(k=0\), and their eigenfrequencies, including dipolar field effects, become
\begin{align}
\tilde{\omega}_{1,2}=\mu_0\gamma l+h_+\mp\sqrt{(\mu_0\gamma H_0)^2+h_-^2},
\label{qu:frequencies_dipolar}
\end{align}
where \(l=\sqrt{(K+J_{AF})^2-J^2_{AF}}\) and the terms \(h_\pm\) arise from dipolar fields:
\begin{align}
h_{\pm}=\frac{1}{2}\mu_0\gamma M_s(1\pm N_{y0})\frac{K}{l}.
\label{qu:h_terms}
\end{align}
Accordingly, even in the absence of the external magnetic field, the two antiferromagnetic modes are no longer degenerate due to the breaking of $PT$ symmetry. 

\subsubsection{Magnon-photon coupling}
\label{qu:coupling}

The coupling between magnons and photons occurs through the Zeeman interaction:
\begin{align}
\hat{H}_{\mathrm{int}}=-\mu_0\int dx\,d{\bm \rho}\,\hat{M}_y({\bm r})\hat{H}_{\mathrm{Sw}}({\bm r}).
\label{qu:Zeeman_interaction}
\end{align}
Since the magnetic field of the Swihart photon \(\hat{H}_{\mathrm{Sw}}\) is along the \(y\)-axis, only the \(y\)-component of the magnetization couples to it. In quantum form, the interaction Hamiltonian takes the form (see Sec.~\ref{sub:coupling} for calculation details):
\begin{align}
\hat{H}_{\mathrm{int}}&=\hbar\sum_{\bm k}\Big(g_1({\bm k})\tilde{m}_{1,\bm k}(\hat{p}_{\bm k}^{\dagger}+\hat{p}_{-\bm k})\nonumber\\
&+g_2({\bm k})\tilde{m}_{2,\bm k}(\hat{p}_{\bm k}^{\dagger}+\hat{p}_{-\bm k})+\mathrm{H.c.}\Big),
\label{qu:ham_interaction}
\end{align}
where \(\hat p_{\bm k}\) are photon operators and \(\tilde m_{1(2),\bm k}\) are magnon operators corresponding to the \(\tilde \omega_{1,2}\) modes, respectively. The coupling constants \(g_{1,2}\) are distinct. According to our calculations in Sec.~\ref{sub:coupling},
\begin{align}
g_1 &\propto \frac{h_-}{\sqrt{h_-^2+\left(\mu_0\gamma H_0-\sqrt{(\mu_0\gamma H_0)^2+h_-^2}\right)^2}}\nonumber\\
&\quad +\frac{h_-}{\sqrt{h_-^2+\left(\mu_0\gamma H_0+\sqrt{(\mu_0\gamma H_0)^2+h_-^2}\right)^2}}, \nonumber \\
g_2 &\propto \frac{\mu_0\gamma H_0-\sqrt{(\mu_0\gamma H_0)^2+h_-^2}}{\sqrt{h_-^2+\left(\mu_0\gamma H_0-\sqrt{(\mu_0\gamma H_0)^2+h_-^2}\right)^2}}\nonumber\\
&\quad +\frac{\mu_0\gamma H_0+\sqrt{(\mu_0\gamma H_0)^2+h_-^2}}{\sqrt{h_-^2+\left(\mu_0\gamma H_0+\sqrt{(\mu_0\gamma H_0)^2+h_-^2}\right)^2}}.
\label{qu:couplings_selectivity}
\end{align}

\subsection{Magnetic-field-controlled coupling selectivity} 
\label{qu:coupling_constants}

A distinctive feature of the S/AF/S system is the tunable selectivity of magnon-photon coupling by an external magnetic field \(\mathbf{H}_0 = H_0 \mathbf{e}_z\), as suggested by Eqs.~\eqref{qu:couplings_selectivity}.

At zero field (\(H_0=0\)), the degeneracy of the two bulk AF magnon modes (see Eq.~\eqref{qu:eigen_bare}) is lifted by the dipolar interaction (Eq.~\eqref{qu:frequencies_dipolar}). The resulting modes are nearly linearly polarized along the \(x\) and \(y\) axes, respectively, see Fig.~\ref{fig:qualitative}(b). Only the mode with finite net magnetization along \(y\) couples to the Swihart photon, forming a bright magnon-polariton with coupling constant \(g_1 \neq 0\). It is seen as upper and lower magnon-polariton branches, $\Omega_2$ and $\Omega_3$, respectively, in Fig.~\ref{fig:qualitative}(c). The mode polarized along \(x\) remains a dark mode, decoupled from the photon, consistent with \(g_2 = 0\). It corresponds to the $\Omega_1$-mode in Fig.~\ref{fig:qualitative}(c). At \(H_0=0\), the bright mode ($\tilde {\omega}_1$) corresponds to sublattices precessing with opposite phases, producing a net \(y\)-magnetization, as shown in Fig.~\ref{fig:qualitative}(b). The dark mode ($\tilde {\omega}_2$) involves precessions that cancel the net \(y\)-component. This intuitive classical picture is detailed in Sec.~\ref{sub:polarization}.

Under a finite magnetic field, the AF eigenmodes evolve from linear toward circular via elliptical polarization. Consequently, both magnon modes develop a finite \(y\)-component of magnetization and couple to the Swihart photon, see Fig.~\ref{fig:qualitative}(b). The coupling strength for each mode varies smoothly with \(H_0\), becoming equal only in the high-field limit (\(\gamma\mu_0 H_0 \gg h_-\)), as seen in Eqs.~\eqref{qu:couplings_selectivity}. This continuous tunability results from fully accounting for dipolar fields and Meissner screening, distinguishing our results from earlier works that neglected dipolar fields, where coupling was essentially binary: any nonzero field was effectively ``strong" and gave equal coupling to both antiferromagnetic modes, with decoupling occurring only at \(H_0 = 0\) \cite{Yuan2017}. The field scale \(h_-\) governing this transition emerges from the dipolar interaction renormalized by the superconductors.

\subsection{Ultrastrong coupling and high cooperativity}
\label{qu:ultra_strong}

For typical parameters (e.g., using MnF\(_2\) as the prototypical AF and NbN as the superconductor), the coupling constant reaches \(|g_1| \sim 100\)~GHz. Given the terahertz-scale bare frequencies of AF magnons (\(\tilde{\omega} \sim 1\)~THz), the ratio \(g/\tilde{\omega} \sim 0.1\) places the system at the boundary of the \emph{ultrastrong coupling regime}. This coupling strength substantially exceeds values reported for AF magnon-polaritons in conventional terahertz cavities \cite{Yuan2017,Grishunin2018,Sivarajah2019,Chen2025,Bialek2020,Bialek2021,Metzger2022,Blank2023} and in analogous superconductor/ferromagnet/superconductor heterostructures \cite{Golovchanskiy2021,Golovchanskiy2021_2,Silaev2022,Qiu2024}. Furthermore, the low dissipation in both the superconducting resonator and the high-quality AF insulator leads to an exceptionally high cooperativity \(C = 4g^2/(\kappa_m \kappa_{ph}) \sim 10^7\), where \(\kappa_m\) and \(\kappa_{ph}\) are the magnon and photon decay rates, respectively, promising coherent quantum applications. See Sec.~\ref{sub:classical_dispersion} for further details on the calculation of magnon and photon decay rates and cooperativity.

\subsection{Properties of the hybrid magnon-polaritons.} 
\label{qu:spin}

The hybrid magnon-polariton branches (three in the general case; see Fig.~\ref{fig:qualitative}(c)) inherit unique characteristics from their constituents.

\textbf{Non-integer and \(k\)-dependent spin.} The magnon-polaritons carry a non-integer average spin \(\langle S_z \rangle < \hbar\), which depends on both the external field \(H_0\) and the wavevector \(k\). The \(z\)-component of spin is not conserved because terms in the quantum Hamiltonian originating from the dipolar interaction do not conserve the number of magnons with a definite spin projection onto the \(z\)-axis, as detailed in Sec.~\ref{sub:magnons}. The average spin of bare magnon modes (before hybridization with the Swihart photon) is independent of \(k\) but non-integer in general. However, it becomes \(k\)-dependent due to magnon-photon hybridization via electromagnetic interaction, as shown in Fig.~\ref{fig:qualitative}(c). A striking prediction is the switching of the spin sign for the middle magnon-polariton branch upon varying \(k\). A similar effect of spin non-conservation due to dipolar interactions was demonstrated in Ref.~\cite{Kamra2017}, but only for a ferrimagnet where bare magnon branches can intersect.

\textbf{High group velocity.} In the strong-mixing region where bare magnon and photon dispersions intersect, the group velocities of the magnon-polariton branches can reach a significant fraction of the speed of light (\(v_g \lesssim c/4\)). This vastly exceeds typical magnon group velocities and enables the prospect of ultrafast and superconductivity-tunable magnon transport in antiferromagnets.

In the following sections, we substantiate this qualitative picture with rigorous derivations. The calculation of modified dipolar fields in the superconducting environment is presented in Sec.~\ref{sec:dipolar}. The quantum theory of magnon-photon coupling, leading to magnon-polariton dispersion and spin, is developed in Sec.~\ref{sec:quantum}. For comparison and physical insight, Sec.~\ref{sec:classical} re-derives the main results using the classical Landau-Lifshitz-Gilbert formalism, explicitly analyzing polarization states. Technical details are deferred to the appendices.

\section{Dipolar fields}
\label{sec:dipolar}

The reconstruction of the magnonic excitations in the antiferromagnet sandwiched between two S layers is caused by the electromagnetic interaction with the dipolar fields, which are strongly modified by the Meissner supercurrents in the superconductors. Those currents are induced by the combination of the Swihart mode of the superconducting resonator and the stray magnonic fields. Consequently, in order to obtain the dispersion relations of the excitations in the system, firstly, we need to calculate the dipolar fields acting on the magnetization, taking into account the superconducting environment of the AF layer.  For this purpose, we solve  Maxwell's equations for the electric and magnetic fields radiated by the magnetization dynamics. Maxwell's equations for the electric field $\bm E(\bm r,t)$, magnetic field $\bm H(\bm r,t)$, magnetic field induction $\bm B(\bm r,t)$, and electric current $\bm J(\bm r,t)$ take the form
\begin{align}
    &\mathrm{rot}~ \bm E=-\dfrac{\partial \bm B}{\partial t},\nonumber\\
    &\mathrm{rot}~ \bm H=\bm J+\varepsilon \dfrac{\partial \bm E}{\partial t},
    \label{Maxwell}
\end{align}
where $\varepsilon$ is the dielectric constant of the corresponding layers. As the electric field is radiated by the magnetization dynamics, it takes the form $\bm E(\bm r,t)=\tilde {\bm E}(x)e^{i\bm k\cdot\bm \rho-i\omega t}$. Considering $\bm B=\mu_0(\bm H+\bm M)$ and $\bm J=\sigma \bm E$, where $\mu_0$ is the vacuum magnetic permeability and $\sigma$ is the conductivity of the layer, from Eqs.~(\ref{Maxwell}) we obtain
\begin{align}
    \Delta \bm E+\mu_0(i\omega\sigma+\varepsilon\omega^2)\bm E+i\omega \mu_0 \mathrm{rot}\bm M=0.
    \label{Maxwell_1}
\end{align}

In the framework of the two-fluid model, the conductivity of a superconductor at frequency $\omega$ takes the form \cite{Schmidt_book}
\begin{align}
\sigma_{S,L} (\omega) = \frac{\rho_n e^2 \tau}{m_e(1+\omega^2 \tau^2)}(1+i\omega \tau) + i \dfrac{\rho_s e^2 }{m_e \omega},  
\label{conductivity_S}
\end{align}
where $\rho_s = \rho_0(1-(T/T_c)^4)$ and $\rho_n = \rho_0 (T/T_c)^4$ are superfluid and normal fluid densities, respectively. Equation~(\ref{conductivity_S}) can then be rewritten as
\begin{align}
\sigma_{S,L} (\omega) = \sigma_n \left(\frac{T}{T_c}\right)^4 (1+i\omega \tau)+ \frac{i}{\omega \mu_0 \lambda_L^2},
    \label{conductivity_S_1}
\end{align}
where  $\sigma_n = \rho_0 e^2 \tau/m_e$ is the conductivity of normal metals, $\lambda_L = \sqrt{m_e/(\mu_0 \rho_s e^2)}$ is London’s
    penetration depth, $\rho_0$ is the electron density, $m_e$ is the electron mass, $e$ is the electron charge, $T_c$ is the critical temperature of the superconductor, $T$ is the temperature of the system, and $\tau$ is the relaxation time of electrons. 

However, under a high-frequency ac field, the relationship between the current and the vector potential changes, which depends in an essential way on the frequency \cite{Abrikosov1958,Abrikosov1960,Mattis1958,ZIMMERMANN199199}. Since in the system with an antiferromagnet we deal with high magnon frequencies $\omega \sim 1$ THz,  the conductivity of the superconductor needs to be calculated using the formalism of microscopic Green's functions. For isotropic BCS superconductors, which is the case under our consideration, we use the expression obtained in Ref.~\cite{ZIMMERMANN199199}:
\begin{align}
    \sigma_S (\omega)=i\dfrac{\sigma_n}{2\omega\tau}I.
    \label{cond}
    \end{align}
Here $I$ is a dimensionless integral over quasiparticle energy, the explicit expression for which can be found in Ref.~\cite{ZIMMERMANN199199}.

In this work, we shall focus on the case $T=0$ and $\hbar \omega < 2 \Delta$, where $\Delta$ is the superconducting order parameter. Then the real part of $\sigma_S$ is absent, and the conductivity Eq.~\eqref{cond} is well fitted by the expression
\begin{align}
    \sigma_S (\omega)=\frac{i}{\omega \mu_0 \lambda_{\rm{eff}}^2}, 
    \label{conductivity_lambda_eff}
\end{align}
where $\lambda_{\rm{eff}}$ is an effective penetration depth, which does not depend on $\omega$.

On the other hand, after substituting explicit expressions for $\lambda_L$ and $\sigma_n$ Eq.~\eqref{conductivity_S_1} is transformed to
\begin{align}
   \sigma_{S,L}(\omega)=\frac{i}{\omega \mu_0 \lambda_L^2}=i\dfrac{\sigma_n}{\omega\tau}.
    \label{conductivity_S_2}
\end{align}
From Eqs.~\eqref{cond}-\eqref{conductivity_S_2}, one obtains the effective  penetration depth $\lambda_{\rm{eff}}$ in terms of $\lambda_L$ and $I$:
\begin{align}
    \lambda_{\rm{eff}}=\sqrt{\frac{2}{I}}\lambda_L.
    \label{lambda_eff}
\end{align}
In all the calculations we assume that the superconductors are NbN; then for numerical estimations one can take $T_c\approx16$ K, $\lambda(T=0)\approx 80$ nm, $\sigma_n=(1\div5)\cdot 10^{5}~\Omega^{-1}\cdot\rm{m}^{-1}$, and $\tau\sim10^{-15}$ s. For these parameters Eqs.~(\ref{cond}) and (\ref{lambda_eff}) at $\omega\leq2\Delta$, which is the case for $\omega \sim 1$ THz, give $I\sim(0.2\div1)\cdot10^{-2}$ and $\lambda_{\rm{eff}}\sim(1\div3)~\mu$m.

Then we can express Eq.~(\ref{Maxwell_1}) in the S and AF layers:
\begin{align}
    &\mathrm{in ~S:}~~\Delta \bm E+k_S^2 \bm E=0,~~k_S=\sqrt{\omega^2 \mu_0\varepsilon_0-\dfrac{1}{\lambda_{\rm{eff}}^2}},\nonumber\\
     &\mathrm{in ~AF:}~~\Delta \bm E+k_{AF}^2 \bm E=-i\omega \mu_0 \mathrm{rot}\bm M,~~k_{AF}=
\omega\sqrt{\mu_0\varepsilon_{AF}},
    \label{Maxwell_final}
\end{align}
where $\varepsilon_0$ and $\varepsilon_{AF}$ are the dielectric constants of vacuum and the antiferromagnet, respectively. Solving Eq. (\ref{Maxwell_final}), we obtain the electric field amplitudes $\tilde {\bm E}(x)$ in each layer:
\begin{align}
    &\mathrm{in ~S1:}~~\tilde {\bm E}(x)=\bm S_1 e^{iB_S(x-d_{AF})},\nonumber\\
     &\mathrm{in ~S2:}~~\tilde {\bm E}(x)=\bm S_2 e^{-iB_S(x+d_{AF})},\nonumber\\
     &\mathrm{in ~AF:}~~\tilde {\bm E}(x)=\bm A^+e^{iAx}+\bm A^- e^{-iAx}+\dfrac{\omega\mu_0}{A^2}\bm k\times\tilde {\bm M},
\label{Maxwell_solutions}
\end{align}
where $A=\sqrt{k_{AF}^2-k^2}$, $B_S=\sqrt{k_S^2-k^2}$, and $\bm S_{1(2)}=(S_{1(2)x}, S_{1(2)y}, S_{1(2)z})^T$ and $\bm A^{\pm}=(A^{\pm}_{x}, A^{\pm}_{y}, A^{\pm}_{z})^T$ are vectors of unknown coefficients. Here we have used the assumption that the AF is thin enough, such that the magnetizations $\bm M_{1,2}$ can be considered as constant along the film normal $x$-axis throughout $2 d_{AF}$. Also, for the considered parameters $B_S\approx i/\lambda_{\rm{eff}}$, which will be substituted in all the expressions containing $B_S$ in the following text. 

From the first equation in (\ref{Maxwell}) we can express the magnetic field components via $\bm E$ and $\bm M$ according to 
\begin{align}
    &H_x=\dfrac{k_yE_z-k_zE_y}{\omega\mu_0}-M_x,\nonumber\\
     &H_y=\dfrac{ik_zE_x-\partial_xE_z}{i\omega\mu_0}-M_y,\nonumber\\
     &H_z=\dfrac{\partial_xE_y-ik_yE_x}{i\omega\mu_0}.
     \label{H_comp}
\end{align}
The $x$-component of the second equation in (\ref{Maxwell}) gives us 
\begin{align}
    E_x=\dfrac{i(k_yH_z-k_zH_y)}{\sigma-i\varepsilon\omega}.
    \label{E_x}
\end{align}

In the S layers we have $|k_ik_j/i\omega\mu_0 (\sigma_S-i\varepsilon\omega)|=|k_ik_j/k_S^2| \sim (k\lambda_{\rm{eff}})^2 \ll 1$ for $i,j\in\{y,z\}$. Substituting Eq.~(\ref{E_x}) into Eq.~(\ref{H_comp}) and taking into account the smallness of the above parameter, we obtain for $H_y$ and $H_z$ in different layers:
\begin{align}
    &\mathrm{in ~S:}~~H_y=-\dfrac{\partial_xE_z}{i\omega\mu_0}, ~~H_z=\dfrac{\partial_xE_y}{i\omega\mu_0},\nonumber\\
    &\mathrm{in ~AF:} ~~\left(
\begin{array}{cc}
H_y\\H_z
\end{array}
\right)=\dfrac{1}{A^2}\hat K_{AF}\left(\begin{array}{cc}
-\left(\dfrac{\partial_xE_z}{i\omega\mu_0}+M_y\right)\\\dfrac{\partial_xE_y}{i\omega\mu_0}
\end{array}
\right),
    \label{H_final}
\end{align}
where
\begin{align}
\hat K_{AF}=\left(
\begin{array}{cc}
\beta & -\dfrac{K_1}{2} \\  -\dfrac{K_1}{2}
 & \alpha
\end{array}
\right),
\end{align}
in which $\alpha=k_{AF}^2-k^2\cos^2\theta$, $\beta=k_{AF}^2-k^2\sin^2\theta$, and $K_1=k^2\sin2\theta$.

Maxwell's equations (\ref{Maxwell}) are accompanied by the boundary conditions, which are the continuity of the field components $E_y, E_z, H_y$, and $H_z$ at the S/AF interfaces $x=\pm d_{AF}$. Implementing those boundary conditions, we can express all the fields via the magnon amplitudes $\tilde {\bm M}$. As for the parameters under consideration $Ad_{AF}\ll 1$, we obtain the following expression for the demagnetization field produced by the magnetic excitations in the antiferromagnet (the details of the derivation are presented in Appendix~A): 
 \begin{align}
     \hat H=-\hat N \hat { M}, ~~\hat H=\left( \begin{array}{c}
          \tilde H_{x}  \\
          \tilde H_{y}
     \end{array}\right),~~\hat {M}=\left( \begin{array}{c}
          \tilde M_{x}  \\
          \tilde M_{y}
     \end{array}\right).
     \label{H_M}
 \end{align}
Here we have introduced the magnetic field amplitudes $\tilde H_{x,y}$ analogously to the amplitudes of the magnetizations and the electric field: $\bm H(\bm r,t)=\tilde {\bm H}(x)e^{i\bm k\cdot\bm \rho-i\omega t}$, and $\hat N$ is the demagnetization tensor. Up to the zero order with respect to $Ad_{AF}$ and $ k\lambda_{\rm{eff}}$ the demagnetization tensor is
\begin{align}
     \hat N&=\left(\begin{array}{cc}
       N_x  & 0 \\
         0 & N_y
     \end{array}\right),\nonumber\\
     N_x&=1,\nonumber\\
     N_y&=\dfrac{d_{AF}\left(d_{AF}k^2\sin^2\theta-k_{AF}^2(d_{AF}+\lambda_{\rm{eff}})\right)}{(d_{AF}+\lambda_{\rm{eff}})\left(d_{AF}k^2-k_{AF}^2(d_{AF}+\lambda_{\rm{eff}})\right)}.
     \label{N}
 \end{align}
This dipolar field is responsible for the formation of magnon-polaritons with ultrastrong coupling between magnon and photon, which we address below by the quantum (Sec. \ref{sec:quantum}) and classical (Sec. \ref{sec:classical}) views, respectively.

\section{Magnon-polaritons: quantum view}
\label{sec:quantum}

Now our goal is to derive the eigenfrequencies of the collective magnon-polariton modes in the S/AF/S heterostructures. In this section, we perform this derivation using quantum formalism.

\subsection{Swihart photon}
\label{sub:swihart}

Physically, the magnons of the antiferromagnet couple with the photon modes of the superconducting resonator. 
To this end, in the total Hamiltonian of the system, we first address the part associated with the photonic Swihart mode in the S/I/S resonator~\cite{Swihart1961}. As was shown in Ref.~\cite{Qiu2024}, the magnetic field $\hat{\bm H}_{Sw}=\hat{H}_{Sw,y}\bm e_y+\hat{H}_{Sw,z}\bm e_z$ for the Swihart photons is quantized according to 
\begin{align}
     \hat{H}_{ Sw,y}&=\sum_{\bm k} \left(\frac{\Omega_s}{2k}\sqrt{\frac{\varepsilon_{AF}\hbar\Omega_s}{d_{AF}}}\cos\theta e^{i{\bm k}\cdot{\bm \rho}}\hat{p}_{\bm k}+{\rm H.c.}\right),\nonumber\\
     \hat{H}_{ Sw,z}&=\sum_{\bm k} \left(-\frac{\Omega_s}{2k}\sqrt{\frac{\varepsilon_{AF}\hbar\Omega_s}{d_{AF}}}\sin\theta e^{i{\bm k}\cdot{\bm \rho}}\hat{p}_{\bm k}+{\rm H.c.}\right),
     \label{magnetic_field_quantization}
\end{align} 
where 
\begin{align}
\Omega_s({\bm k})=\sqrt{\frac{d_{AF}}{\mu_0\varepsilon_{AF}(d_{AF}+\lambda_{\rm{eff}})}}|{\bm k}|
\label{Swihart_mode}
\end{align} 
is the dispersion relation of the Swihart mode, and $\{\hat{p}^{\dagger}_{\bm k},\hat{p}_{\bm k}\}$ are the photon creation and annihilation operators that obey the commutation relation $[\hat{p}_{\bm k}, \hat{p}^{\dagger}_{{\bm k}'} ]=i\hbar\delta_{{\bm k}{\bm k}'}$. The Hamiltonian of the Swihart mode then reads 
\begin{align}
\hat H_{ph}=\hbar\sum_{\bm k}\Omega_s({\bm k})\hat{p}_{\bm k}^{\dagger}\hat{p}_{\bm k}.
\label{Swihart_H}
\end{align} 

\subsection{Uncoupled magnons}
\label{sub:magnons}

In the second step, let us study the Hamiltonian describing magnons in the antiferromagnet at $k \to 0$, when the Swihart mode is absent. We introduce the magnetization operators $\hat{\bm M}_{1,2}({\bm r})$ for the magnetizations of the two sublattices. The magnetizations $\hat{\bm M}_{1,2}({\bm r})$ are affected by the effective magnetic field, including the in-plane static external magnetic field, along the N\'eel vector ${\bm z}$-direction
\begin{align}
    \bm H_0({\bm r})=H_0 \bm e_z,
    \label{h0}
\end{align}
the exchange magnetic field
\begin{align}
    {\bm H}^{\rm ex}_{1,2}({\bm r})=-\frac{J_{ AF}}{M_s} \bm M_{2,1},
    \label{Hex}
\end{align}
the anisotropy magnetic field
\begin{align}
    {\bm H}^{\rm ani}_{1,2}({\bm r})=\frac{K}{M_s}\hat{M}_{1,2z}({\bm r})\bm e_z,
    \label{H_ani}
\end{align}
and the demagnetization field
\begin{align}
    {\bm H}^d({\bm r})= -N_{x0}\left(\hat{M}_{1x}+\hat{M}_{2x}\right)\bm e_x-N_{y0}\left(\hat{M}_{1y}+\hat{M}_{2y}\right)\bm e_ y.
    \label{H_d}
\end{align}
Here $K$ and $J_{AF}$ are the anisotropy and exchange constants, respectively, and $N_{x(y)0} \equiv N_{x(y)}(k=0)$ are the components of the demagnetization tensor (\ref{N}) taken at $k=0$. With these effective magnetic fields, the Hamiltonian of the antiferromagnetic magnons takes the form
\begin{align}
    &\hat{H}_{\rm AFM}=\sum_{i=1,2}\int d{\bm r} \Big\{-\mu_0\hat{\bm M}_i({\bm r})\cdot {\bm H}_{0}({\bm r}) \nonumber \\
    &-\frac{\mu_0}{2}\hat{\bm M}_i({\bm r})\cdot \left[{\bm H}^{\rm ex}_{i}({\bm r})+{\bm H}^d({\bm r})+{\bm H}^{\rm ani}_{i}({\bm r})\right]\Big\}.
    \label{H_m}
\end{align}

Then we introduce the operators of the spin density $\hat{\bm S}_i({\bm r})$ at the sublattice $i$, so that $\hat{\bm M}_i({\bm r})=-\gamma \hbar \hat{\bm S}_i({\bm r})$ and $M_s=\gamma \hbar S$, where $S$ is the total spin of the system. As the next step, we represent the spin-density operators $\hat{\bm S}_i({\bm r})$ in terms of the bosonic operators $\hat{a}_i({\bm r})$ via the
Holstein-Primakoff transformation \cite{HolsteinPrimakoff}:
    \begin{align}
    \hat{S}_{1x}&=\frac{\sqrt{2S}}{2}\left(\hat{a}^{\dagger}_1+\hat{a}_1\right),~~~ \hat{S}_{2x}=\frac{\sqrt{2S}}{2}\left(\hat{a}^{\dagger}_2+\hat{a}_2\right), \nonumber\\
    \hat{S}_{1y}&=\frac{\sqrt{2S}}{2i}\left(\hat{a}^{\dagger}_1-\hat{a}_1\right),~~~ \hat{S}_{2y}=\frac{\sqrt{2S}}{2i}\left(\hat{a}_2-\hat{a}^{\dagger}_2\right), \nonumber\\
    \hat{S}_{1z}&=-S+\hat{a}^{\dagger}_1\hat{a}_1,~~~~~~~~~~ \hat{S}_{2z}=S-\hat{a}^{\dagger}_2\hat{a}_2. 
\label{HP}
\end{align}
When the wave vector ${\bm k}$ is small, we focus on the modes close to the antiferromagnetic resonance, which are uniform across the antiferromagnetic film, i.e., there is no $x$-dependence. After the Fourier transformation $\hat{a}_i(-d_{AF}\le x\le d_{AF},{\bm \rho})=(1/\sqrt{2d_{AF}})\sum_{\bm k}\hat{a}_i({\bm k})e^{i{\bm k}\cdot {\bm \rho}}$, Eq.~(\ref{HP}) can be rewritten as ($-d_{AF}\le x\le d_{AF}$):
\begin{widetext}
\begin{align}
    \hat{S}_{1x}({\bm \rho})&=\frac{\sqrt{2S}}{2\sqrt{2d_{AF}}}\sum_{\bm k}\left(\hat{a}^{\dagger}_{1,{\bm k}}e^{-i{\bm k}\cdot{\bm \rho}}+\hat{a}_{1,{\bm k}}e^{i{\bm k}\cdot{\bm \rho}}\right), ~~~~\hat{S}_{2x}({\bm \rho})=\frac{\sqrt{2S}}{2\sqrt{2d_{AF}}}\sum_{\bm k}\left(\hat{a}^{\dagger}_{2,{\bm k}}e^{-i{\bm k}\cdot{\bm \rho}}+\hat{a}_{2,{\bm k}}e^{i{\bm k}\cdot{\bm \rho}}\right),
    \nonumber\\
    \hat{S}_{1y}({\bm \rho})&=\frac{\sqrt{2S}}{2i\sqrt{2d_{AF}}}\sum_{\bm k}\left(\hat{a}^{\dagger}_{1,{\bm k}}e^{-i{\bm k}\cdot{\bm \rho}}-\hat{a}_{1,{\bm k}}e^{i{\bm k}\cdot{\bm \rho}}\right), ~~~\hat{S}_{2y}({\bm \rho})=\frac{\sqrt{2S}}{2i\sqrt{2d_{AF}}}\sum_{\bm k}\left(\hat{a}_{2,{\bm k}}e^{i{\bm k}\cdot{\bm \rho}}-\hat{a}^{\dagger}_{2,{\bm k}}e^{-i{\bm k}\cdot{\bm \rho}}\right), \nonumber\\
    \hat{S}_{1z}({\bm \rho})&=-S+\frac{1}{2d_{AF}}\sum_{{\bm k}{\bm k}'}\hat{a}^{\dagger}_{1,{\bm k}}\hat{a}_{1,{\bm k}'}e^{i({\bm k}'-{\bm k})\cdot{\bm \rho}}, ~~~~~~~~~\hat{S}_{2z}({\bm \rho})=S-\frac{1}{2d_{AF}}\sum_{{\bm k}{\bm k}'}\hat{a}^{\dagger}_{2,{\bm k}}\hat{a}_{2,{\bm k}'}e^{i({\bm k}'-{\bm k})\cdot{\bm \rho}}. 
    \label{HP2}
\end{align}
Substituting Eq. (\ref{HP2}) into Eq.  (\ref{H_m}), we can present the magnon Hamiltonian in the following matrix form:
\begin{align}
&\hat{H}_{\rm AFM}=\sum_{\bm k}\hat{X}^{\dagger}({\bm k}){\cal H}_{0}\hat{X}({\bm k}), ~~~~\hat{X}({\bm k})=\begin{pmatrix}\hat{a}_{1,{\bm k}}\\\hat{a}_{2,{\bm k}}\\\hat{a}^{\dagger}_{1,-{\bm k}}\\\hat{a}^{\dagger}_{2,-{\bm k}}\end{pmatrix},\nonumber\\
    &{\cal H}_{0}=\mu_0 \gamma \hbar    \begin{pmatrix}
        F_++\dfrac{M_s(N_{x0}+N_{y0})}{2} & \dfrac{M_s(N_{x0}-N_{y0})}{2} & \dfrac{M_s(N_{x0}-N_{y0})}{2} & J_{ AF}+\dfrac{M_s(N_{x0}+N_{y0})}{2}) \\
        \dfrac{M_s(N_{x0}-N_{y0})}{2} & F_-+\dfrac{M_s(N_{x0}+N_{y0})}{2}& J_{ AF}+\dfrac{M_s(N_{x0}+N_{y0})}{2} & \dfrac{M_s(N_{x0}-N_{y0})}{2} \\
        \dfrac{M_s(N_{x0}-N_{y0})}{2}& J_{ AF}+\dfrac{M_s(N_{x0}+N_{y0})}{2}& F_++\dfrac{M_s(N_{x0}+N_{y0})}{2} & \dfrac{M_s(N_{x0}-N_{y0})}{2} \\
        J_{ AF}+\dfrac{M_s(N_{x0}+N_{y0})}{2}& \dfrac{M_s(N_{x0}-N_{y0})}{2} & \dfrac{M_s(N_{x0}-N_{y0})}{2} & F_-+\dfrac{M_s(N_{x0}+N_{y0})}{2}
    \end{pmatrix},
    \label{H_matrix}
\end{align}
\end{widetext}
where $F_{\pm}=K+J_{AF}\pm H_0$.

When disregarding the demagnetization effect by taking $N_{x0}=N_{y0}=0$, the eigenfrequencies of the Hamiltonian (\ref{H_matrix}) take the form
\begin{align}
\hbar \omega_{1,2}^b=\mu_0\gamma\hbar \left(\sqrt{(K+J_{ AF})^2-J^2_{ AF}}\pm H_0\right), 
\label{eigen_bare}
\end{align}
which represent widely known eigenmodes of the antiferromagnetic resonance for the case of an easy-axis antiferromagnet \cite{Rezende2019}. The transformation matrix ${\cal T}$, which diagonalizes ${\cal H}_{0}$ at $N_{x0}=N_{y0}=0$, is introduced as
\begin{align}
    \hat{X}({\bm k})={\cal T} \hat{\Psi}({\bm k}),
    \label{a-m}
\end{align}
where $\hat{\Psi}({\bf k})=(\hat{m}_{1,{\bf k}},\hat{m}_{2,{\bf k}}, \hat{m}^{\dagger}_{1,-{\bf k}},\hat{m}^{\dagger}_{2,-{\bf k}})^T$ is a vector of new magnon operators $\hat{m}_{1(2),{\bm k}}$. We adopt a hyperbolic parametrization in terms of the parameters $\{\varphi,l\}$: $K+J_{ AF}=l\cosh{\varphi}$ and $J_{ AF}=l\sinh{\varphi}$, such that $\hbar \omega_{1,2}^b=\mu_0\gamma\hbar(l\pm H_0)$. Then the corresponding  Bogoliubov transformation matrix is 
\begin{align}
    {\cal T}=\begin{pmatrix}
        -\cosh{\frac{\varphi}{2}} &0&0& -\sinh{\frac{\varphi}{2}} \\0&\cosh{\frac{\varphi}{2}}&\sinh{\frac{\varphi}{2}}&0\\0&-\sinh{\frac{\varphi}{2}} &-\cosh{\frac{\varphi}{2}} &0\\
        \sinh{\frac{\varphi}{2}}  & 0&0& \cosh{\frac{\varphi}{2}}        
    \end{pmatrix},
    \label{Tmatrix}
\end{align} 
where $\cosh ({\varphi}/{2})=\sqrt{({K+J_{ AF}+l})/({2l})}$, $\sinh ({\varphi}/{2})=\sqrt{({K+J_{AF}-l})/({2l})}$, and $l=\sqrt{(K+J_{ AF})^2-J^2_{AF}}$. With the use of Eqs.~(\ref{HP2}), (\ref{a-m}), and (\ref{Tmatrix}), the magnetization operators are quantized as ($-d_{AF}\le x\le d_{AF}$)
\begin{align}
    \hat{M}_{1x}({\bm \rho})&=\sqrt{\frac{\gamma \hbar M_s}{4d_{AF}}}\sum_{\bm k}\Big(\cosh{\frac{\varphi}{2}}\hat{m}_{1,{\bm k}}e^{i{\bm k}\cdot{\bm \rho}} \nonumber \\
    &+\sinh{\frac{\varphi}{2}}\hat{m}_{2,{\bm k}}e^{i{\bm k}\cdot{\bm \rho}}+{\rm H.c.}\Big), \nonumber\\
    \hat{M}_{1y}({\bm \rho})&=-i\sqrt{\frac{\gamma \hbar M_s}{4d_{AF}}}\sum_{\bm k}\Big(-\cosh{\frac{\varphi}{2}}\hat{m}_{1,{\bm k}}e^{i{\bm k}\cdot{\bm \rho}} \nonumber\\
    &+\sinh{\frac{\varphi}{2}}\hat{m}_{2,{\bm k}}e^{i{\bm k}\cdot{\bm \rho}}-{\rm H.c.}\Big), \nonumber \\
    \hat{M}_{2x}({\bm \rho})&=-\sqrt{\frac{\gamma \hbar M_s}{4d_{AF}}}\sum_{\bm k}\Big(\sinh{\frac{\varphi}{2}}\hat{m}_{1,{\bm k}}e^{i{\bm k}\cdot{\bm \rho}} \nonumber \\
    &+\cosh{\frac{\varphi}{2}}\hat{m}_{2,{\bm k}}e^{i{\bm k}\cdot{\bm \rho}}+{\rm H.c.}\Big), \nonumber\\
    \hat{M}_{2y}({\bm \rho})&=i\sqrt{\frac{\gamma \hbar M_s}{4d_{AF}}}\sum_{\bm k}\Big(-\sinh{\frac{\varphi}{2}}\hat{m}_{1,{\bf k}}e^{i{\bf k}\cdot{\pmb \rho}} \nonumber \\
    &+\cosh{\frac{\varphi}{2}}\hat{m}_{2,{\bf k}}e^{i{\bf k}\cdot{\pmb \rho}}-{\rm H.c.}\Big). 
    \label{magnetization_quantization}
\end{align}
The magnetization fluctuation along the $y$-direction then reads
\begin{align}
    &\hat{M}_y({\bm \rho})=\hat{M}_{1y}({\bm \rho})+\hat{M}_{2y}({\bm \rho})
    =i\sqrt{\frac{\gamma \hbar M_s}{4d_{AF}}} \nonumber \\
    &\times\sum_{\bm k}\left[\left(\cosh\frac{\varphi}{2}-\sinh\frac{\varphi}{2}\right)e^{i{\bm k}\cdot {\bm \rho}}(\hat{m}_{1,{\bm k}}+\hat{m}_{2,{\bm k}})-{\rm H.c.}\right].
    \label{M_y_1}
\end{align}

However, for the considered case of the antiferromagnet sandwiched between two S layers, the demagnetization effect produced by the S layers is not small as compared to the effects of anisotropy and external field and, therefore, cannot be neglected. Thus, taking into account the demagnetization field and
by substituting Eq.~\eqref{magnetization_quantization} into Eq.~\eqref{H_m}, we find the coupled magnon Hamiltonian
\begin{align}
 \hat{H}_m&=\hbar\sum_{\bm k}\bigg[(\omega_1+h_+)\hat{m}^{\dagger}_{1,{\bm k}}\hat{m}_{1,{\bm k}}+(\omega_2+h_+)\hat{m}^{\dagger}_{2,{\bm k}}\hat{m}_{2,{\bm k}} \nonumber \\
 &-h_-\left(\hat{m}_{1,{\bm k}}\hat{m}^{\dagger}_{2,{\bm k}}+{\rm H.c.}\right)-h_+\left(\hat{m}_{1,{\bm k}}\hat{m}_{2,-{\bm k}}+{\rm H.c.}\right) \nonumber \\
 &+\frac{1}{2}h_-\left(\hat{m}_{1,{\bm k}}\hat{m}_{1,-{\bm k}}+\hat{m}_{2,{\bm k}}\hat{m}_{2,-{\bm k}}+ {\rm H.c.}\right)\bigg]
 \label{ham_m}
\end{align}
with the coupling constants 
\begin{align}
    h_{\pm}=\frac{1}{2}\mu_0\gamma M_s(N_{x0}\pm N_{y0})\left(\cosh{\varphi}-\sinh{\varphi}\right).
\end{align}
After the diagonalization procedure, we obtain the following dispersion relations of the two magnon modes (before the spin-flop transition): 
\begin{align}
    \omega_{1,2}=\mu_0\gamma\sqrt{H^2_0+H_b^2\pm\sqrt{K^2M_s^2(N_{x0}-N_{y0})^2+4H_0^2H_b^2}},
    \label{E_k}
\end{align}
in which $H_{b}^2 = 2J_{AF}K+K^2+KM_s(N_{x0}+N_{y0})$.
Since analytical solutions for the eigenstates are too lengthy to be listed,  we then seek an approximate solution by treating the demagnetization effect as a perturbation.
Since $h_{\pm} \ll \omega_{1,2}$, we disregard the squeezing terms and arrive at 
\begin{align}
 \hat{H}_m &\approx \hbar\sum_{\bm k}\bigg[(\omega_1+h_+)\hat{m}^{\dagger}_{1,{\bm k}}\hat{m}_{1,{\bm k}}+ (\omega_2+h_+)\hat{m}^{\dagger}_{2,{\bm k}}\hat{m}_{2,{\bm k}}\nonumber \\
 &- h_-\left(\hat{m}_{1,{\bm k}}\hat{m}^{\dagger}_{2,{\bm k}}+\hat{m}^{\dagger}_{1,{\bm k}}\hat{m}_{2,{\bm k}}\right)\bigg].
 \label{H_m1}
\end{align}
Next, we diagonalize the Hamiltonian Eq.~\eqref{H_m1} 
by introducing the new operators of magnon modes $\tilde{m}_{1(2),\bm k}$:
\begin{align}
    \tilde{m}_{1,\bm k}=\frac{h_-}{\sqrt{h_-^2+\left(\mu_0\gamma H_0-\sqrt{(\mu_0\gamma H_0)^2+h_-^2}\right)^2}}\hat{m}_{1,{\bm k}} \nonumber \\
    +\frac{h_-}{\sqrt{h_-^2+\left(\mu_0\gamma H_0+\sqrt{(\mu_0\gamma H_0)^2+h_-^2}\right)^2}}\hat{m}_{2,{\bm k}},\nonumber\\
    \tilde{m}_{2,\bm k}=\frac{\mu_0\gamma H_0-\sqrt{(\mu_0\gamma H_0)^2+h_-^2}}{\sqrt{h_-^2+\left(\mu_0\gamma H_0-\sqrt{(\mu_0\gamma H_0)^2+h_-^2}\right)^2}}\hat{m}_{1,{\bm k}} \nonumber \\
    +\frac{\mu_0\gamma H_0+\sqrt{(\mu_0\gamma H_0)^2+h_-^2}}{\sqrt{h_-^2 + \left(\mu_0\gamma H_0+\sqrt{(\mu_0\gamma H_0)^2+h_-^2}\right)^2}}\hat{m}_{2,{\bm k}}.
    \label{m_new}
\end{align}
$\tilde{m}_{1(2),\bm k}$ are the operators of magnon excitations that take into account the renormalization of the {\it bare} magnon excitations $\hat {m}_{1(2),{\bm k}}$ by the magnetic fields of the Meissner currents, which, in turn, are induced in the S layers due to the magnonic stray fields. 
The magnon Hamiltonian then takes its final form
\begin{align}
 \hat{H}_m=\hbar\sum_{\bm k}\left(\tilde{\omega}_1\tilde{m}^{\dagger}_{1,\bm k}\tilde{m}_{1,\bm k}+\tilde{\omega}_2 \tilde{m}^{\dagger}_{2,\bm k}\tilde{m}_{2,\bm k}\right),
\end{align}
with the magnon frequencies 
\begin{align}
\tilde{\omega}_{1,2}=\mu_0\gamma l+h_+\mp\sqrt{(\mu_0\gamma H_0)^2+h_-^2}.
\end{align}
In terms of operators $\tilde{m}_{1(2),\bm k}$ the magnetization fluctuation along the $y$-direction is quantized as
\begin{align}
    \hat{M}_y({\bm \rho})&=i\sqrt{\frac{\gamma \hbar M_s}{4d_{AF}}}\sum_{\bm k}\Big[\left(\cosh\frac{\varphi}{2}-\sinh\frac{\varphi}{2}\right)\nonumber\\
    &\times\left[ a\tilde{m}_{1,\bm k}+b\tilde{m}_{2,\bm k}\right]e^{i{\bm k}\cdot {\bm \rho}}-{\rm H.c.}\Big]
    \label{M_y_2}
\end{align}
with
\begin{align}
    a=\frac{h_-}{\sqrt{h_-^2+\left(\mu_0\gamma H_0-\sqrt{(\mu_0\gamma H_0)^2+h_-^2}\right)^2}}\nonumber\\+\frac{h_-}{\sqrt{h_-^2+\left(\mu_0\gamma H_0+\sqrt{(\mu_0\gamma H_0)^2+h_-^2}\right)^2}},\nonumber\\
    b=\frac{\mu_0\gamma H_0-\sqrt{(\mu_0\gamma H_0)^2+h_-^2}}{\sqrt{h_-^2+\left(\mu_0\gamma H_0-\sqrt{(\mu_0\gamma H_0)^2+h_-^2}\right)^2}}\nonumber\\+\frac{\mu_0\gamma H_0+\sqrt{(\mu_0\gamma H_0)^2+h_-^2}}{\sqrt{h_-^2+\left(\mu_0\gamma H_0+\sqrt{(\mu_0\gamma H_0)^2+h_-^2}\right)^2}}.
    \label{ab}
\end{align}

\subsection{Magnon-photon coupling}
\label{sub:coupling}

Next, we write the part of the total Hamiltonian that describes the coupling between magnons and photons through the Zeeman interaction. From Eqs.~\eqref{M_y_2} and \eqref{magnetic_field_quantization} we obtain
\begin{align}
    \hat{H}_{int}&=-\mu_0\int dxd{\bm \rho}\hat{\bm M}({\bm r})\cdot\hat{\bm H}_{Sw}({\bm r})\nonumber\\
    &=-\mu_0\int dxd{\bm \rho}\hat{M}_y({\bm r})\hat{H}_{{ Sw},y}({\bm r})\nonumber\\
    &=\hbar\sum_{\bm k}\Big(g_1({\bm k})\tilde{m}_{1,\bm k}(\hat{p}_{\bm k}^{\dagger}+\hat{p}_{-\bm k})\nonumber\\
    &+g_2({\bm k})\tilde{m}_{2,\bm k}(\hat{p}_{\bm k}^{\dagger}+\hat{p}_{-\bm k})+{\rm H. c.}\Big),
    \label{ham_interaction}
\end{align}
where we have introduced the coupling constants 
\begin{align}
g_1({\bm k})&=-i a \sqrt{\mu_0\gamma M_s}
\left(\cosh\frac{\varphi}{2}-\sinh\frac{\varphi}{2}\right)
\nonumber\\
&\times \frac{\Omega_s(k)}{2k}\sqrt{\mu_0\varepsilon_{AF} \Omega_s(k)}\cos\theta ,\nonumber\\
g_2({\bm k})&=-i b \sqrt{\mu_0\gamma M_s}
\left(\cosh\frac{\varphi}{2}-\sinh\frac{\varphi}{2}\right) \nonumber\\
&\times \frac{\Omega_s(k)}{2k}\sqrt{\mu_0\varepsilon_{AF} \Omega_s(k)}\cos\theta .
\label{coupling}
\end{align}
Since the coupling constants depend on the propagation direction $\theta$ with respect to the N\'eel vector by $\cos\theta$, the coupling between magnons and Swihart photon vanishes in the Damon-Eshbach configuration $\theta=\pi/2$.
From Eq.~\eqref{ab} it can be seen that in the limiting case of small external magnetic field  ($\gamma\mu_0H_0\ll h_-$), $a\approx \sqrt{2}$ and $b\approx 0$, which means that only one magnon mode $\tilde{m}_{1,{\bm k}}$ (the one with a lower frequency) couples with the Swihart photon. On the contrary, when $\gamma\mu_0H_0\gg h_-$, $a\approx 1$ and $b\approx 1$, so the two antiferromagnetic modes couple equally with the Swihart mode. Substituting the expression for the Swihart mode frequency Eq.~(\ref{Swihart_mode}) into Eqs.~(\ref{coupling}) we see that the coupling disappears at $\lambda_{\rm eff} \to \infty$. Formally, this limit corresponds to the normal state of the S layers. However, it is worth mentioning here that our answer is only applicable to the superconducting low-temperature state of the S layers. The physics of the antiferromagnetic THz magnon-photon coupling near the critical temperature and in the normal state of the S layers requires a separate consideration because in this case the penetration depth $\lambda_{\rm eff}$ is replaced by the skin depth and the conductivity of the layers acquires a significant real part, which may give rise to an additional non-Hermitian physics including exceptional points \cite{Qiu2024}.

\subsection{Magnon-polaritons: dispersion}
\label{sub:dispersion}

Finally, let us write the total Hamiltonian of the system:
\begin{align}
    &\hat{H}_{ tot}=\hat{H}_{ m}+\hat{H}_{ ph}+\hat{H}_{ int}\nonumber\\
    &=\hbar\sum_{\bm k}\bigg[ \tilde{\omega}_1\tilde{m}_{1,\bm k}^{\dagger}\tilde{m}_{1,\bm k}+\tilde{\omega}_2\tilde{m}_{2,\bm k}^{\dagger}\tilde{m}_{2,\bm k}+\Omega_s\hat{p}_{\bm k}^{\dagger}\hat{p}_{\bm k}\nonumber\\
    &+\Big\{\left(g_1({\bm k})\tilde{m}_{1,\bm k}+g_2({\bm k})\tilde{m}_{2,\bm k}\right)(\hat{p}_{\bm k}^{\dagger}+\hat{p}_{-\bm k})+{\rm H. c.}\Big\}\bigg].
    \label{H_tot}
\end{align}
By performing the Bogoliubov transformation for $\hat{H}_{tot}$, we obtain the eigenfrequencies of the hybrid magnon-polariton modes. The expressions for the eigenfrequencies in the general case are rather cumbersome and therefore are presented in Appendix B, while in the limiting case $\gamma\mu_0H_0\ll h_-$, we obtain $\tilde{\omega}_{1,2}\approx h_+\mp h_-+\gamma\mu_0l$ and $g_2\approx0$, which leads to the eigenfrequencies
\begin{align}
    &\Omega_1 = h_++h_-+\gamma\mu_0l, \label{Omega_1}\\
    &\Omega_{2,3}(\bm k)=\frac{1}{\sqrt{2}}\sqrt{\tilde{\omega}_1^2+\Omega_s^2\pm\sqrt{(\tilde{\omega}_1^2-\Omega_s^2)^2+16|g_1(\bm k)|^2\Omega_s\tilde{\omega}_1}}.
    \label{Omega_23}
\end{align} 
It can be seen that $\Omega_1 = \tilde \omega_2$, which means this mode is not coupled to the Swihart mode, as was mentioned above. The physical reason is that the coupling is provided by $M_y$-component of the AF magnetization, see Eq.~(\ref{ham_interaction}). According to Eq.~(\ref{M_y_2}), $M_y=0$ for the $\Omega_1$ eigenmode, as $b=0$ and only magnon $\tilde m_{2,\bm k}$ is excited in this mode. At the same time, the other mode is coupled to the photon, and the coupled eigenfrequencies $\Omega_{2,3}$ expressed by Eq.~(\ref{Omega_23}) coincide with the answer for the coupled magnon-polariton modes obtained in Refs.~\cite{Silaev2022,Qiu2024} for the S/F/S heterostructures. The coincidence is not accidental, since the coupling mechanism in our case is the same: the coupling is realized through the electromagnetic interaction between $M_y$-component of the AF magnetization and the magnetic field of the Swihart mode. The situation when only one of two magnonic eigenmodes is coupled to the Swihart mode is also to some extent similar to the case of S/F/I/F/S system \cite{Gordeeva2025} with identical ferromagnets, where magnonic modes of two ferromagnets were coupled through the electromagnetic interaction with the Meissner currents in the S layers, and only one of them - acoustic mode - was coupled to the photon, but the other, optical mode, was not coupled due to the absence of net magnetization, and, thus represents an example of a dark mode.

We shall consider the bulk volume configuration with $\bm k$ aligned parallel to $\bm M_{1s}$, i.e., $\theta=0$. Taking for numerical estimates the parameters typical for $\mathrm{MnF}_2$ with thickness $2d_{AF}\approx\lambda_{\rm{eff}}$,  $\varepsilon_{AF}=5.5 \varepsilon_0$, $\gamma=1.76\cdot10^{11}$ Hz$\cdot\mathrm{T}^{-1}$, $\mu_0M_s=0.75$ T, $\mu_0J_{AF}=50$ T, and $\mu_0K=1$ T, we obtain that the coupling strength  for the S/AF/S heterostructures $|g_1(k_K)| \sim  100$ GHz [$k_K$ is the wave number of the photon at the magnon frequency and is determined from the equation $\tilde \omega_1 = \Omega_s(k_K)$]. According to Eq.~(\ref{coupling}) the coupling strength in the S/AF/S system $|g_1(k_K)| \sim \sqrt[4]{K/J_{AF}} \sqrt{d_{AF}/2(d_{AF}+\lambda_{\rm eff})}\sqrt{\mu_0 \gamma M_s \tilde \omega_1}$, and the coupling strength for the S/F/S heterostructure $g_F =   (1/2)\sqrt{d_{F}/(d_{F}+\lambda)} \mu_0 \gamma \sqrt{M_{sF} (M_{sF}+H_0)}$ \cite{Silaev2022,Qiu2024}, where $M_{sF}$ is the saturation magnetization of the ferromagnet, $2d_F$ is the thickness of the ferromagnet. Their ratio $g_F/|g_1(k_K)|= (1/2)\sqrt{(M_{sF}/M_s)(H_0+M_{sF})/K}$ at the same value of the parameter $\sqrt{d/(d+\lambda)}$. It depends on the particular magnetic parameters of the F and AF materials. In our case, the high value of the coupling constant is due to the high value of the magnetic anisotropy constant $K$. This value is approximately by the order of magnitude larger than it was obtained for the coupling strength of AF magnons to photons in THz cavities \cite{Yuan2017,Kritzell2023,Grishunin2018}.   Despite the high value of the coupling constant, it is by the order of magnitude smaller than the typical AF magnon frequencies, which are in the THz range. Thus, the magnon-polaritons in S/AF/S systems are at the boundary of the ultra-strong coupling regime since $|g_1(k_K)|/\tilde \omega_1 \sim 0.1$.    

In the general case of arbitrary magnitudes of the applied magnetic fields, the eigenfrequencies of the magnon-polaritons are expressed by Eqs.~(\ref{Omega_qv}) and are plotted in Fig.~\ref{fig:eigenfrequencies_qv} for three typical values of the external magnetic field $H_0$: zero external field [Fig.~\ref{fig:eigenfrequencies_qv}(a)], moderate external field
$\gamma\mu_0 H_0 \sim h_-$ [Fig.~\ref{fig:eigenfrequencies_qv}(b)], and strong external field $\gamma\mu_0 H_0 \gg h_-$ [Fig.~\ref{fig:eigenfrequencies_qv}(c)]. It is seen that at nonzero magnetic field both modes are coupled to the Swihart mode according to the fact that both coupling constants $g_1$ and $g_2$ are nonzero in this case. At large magnetic fields $\gamma\mu_0 H_0 \gg h_-$, both magnonic modes are equally coupled to the Swihart mode. 

\subsection{Magnon-polaritons: spin and group velocity}
\label{sub:spin}

In addition to being at the boundary of the ultra-strong coupling, the complex magnon-polariton excitations have other important properties. First of all, they are bosonic quasiparticles with non-integer average spin $\langle S_z \rangle$ smaller than $\hbar$. The average spin of all magnon-polariton excitation branches is shown by the color of the lines in Fig.~\ref{fig:eigenfrequencies_qv}. The details of the calculation of the average spin are presented in the Appendix C.
\begin{figure} 
    \centering
    \includegraphics[width=\linewidth]{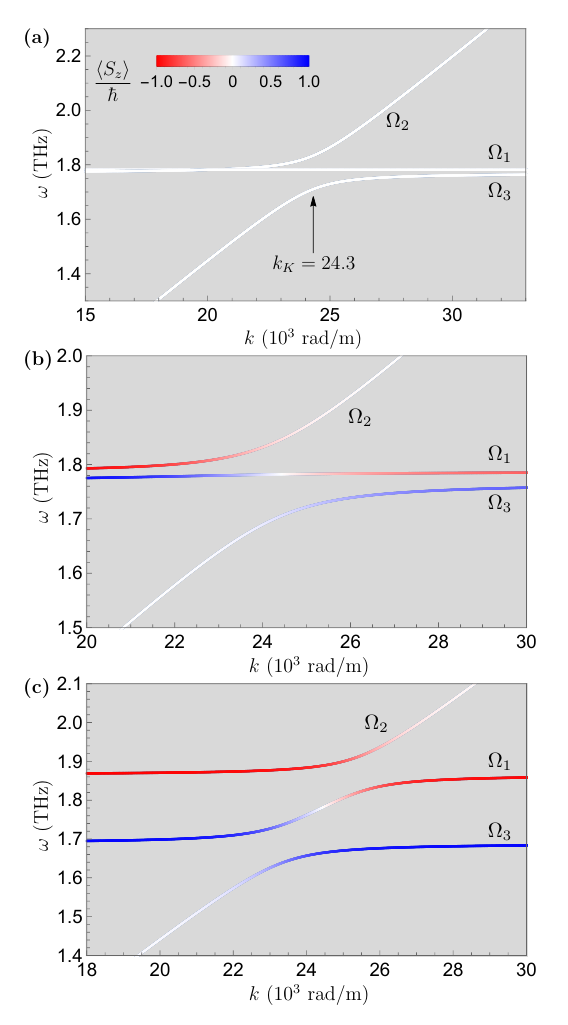}
    \caption{Dispersion of the eigenmodes in the S/AF/S structure in the cases of no external field  [(a)], moderate external field $\mu_0H_0=0.05$ T [(b)], and strong external field $\mu_0H_0=0.5$ T [(c)]. The colors of the curves depict the value of the average spin $\langle S_z \rangle$ on the eigenmodes. Parameters: $2d_{AF}=\lambda_{\rm{eff}}$,  $\varepsilon_{AF}=5.5 \varepsilon_0$, $\gamma=1.76\cdot10^{11}$ Hz$\cdot\mathrm{T}^{-1}$, $\mu_0M_s=0.75$ T, $\mu_0J_{AF}=50$ T,  $\mu_0K=1$ T, and $\theta=0$.}
    \label{fig:eigenfrequencies_qv}
\end{figure}

At $H_0=0$, $\langle S_z \rangle = 0$ for all branches of the spectrum. As it was reported earlier \cite{Kamra2017,Xiyin_1,Xiyin_2}, this is a consequence of mode hybridization taking into account the dipolar interaction, which couples $S_z=\pm \hbar$ magnons described by the operators $\hat m_{1(2), \bm k}$, see Eq.~(\ref{ham_m}). At $H_0 \neq 0$ $\langle S_z \rangle $ is nonzero for all branches and depends on $k$. At $k = 0$ the average spin of the magnonic excitation $\tilde m_{1(2), \bm k}$ grows in absolute value tending to $\hbar (-\hbar)$ at $\gamma\mu_0 H_0 \gg h_-$, which is in agreement with the results for an antiferromagnet taking into account the dipolar interaction \cite{Kamra2017}. At $k \neq 0$ $\langle S_z \rangle $ depends significantly on the wave number $k$ and smoothly changes its value between $S_0$ at $k=0$ and $-S_{\infty}$ at $k \to \infty$ for the middle branch, $0$ and $S_{\infty}$ for the lower branch, and $-S_0$ and $0$ for the upper branch of the spectrum. At $\gamma\mu_0 H_0 \gg h_-$, $S_0 \to \hbar$. Obviously, $\langle S_z \rangle=0 $ at  $k=0$ and $k \to \infty$ for the upper and lower branches is explained by the fact that the magnon part of the excitation is negligibly small, and the maximum absolute value of $S_0 = \hbar$ corresponds to the disappearance of the photonic part and negligible influence of stray fields. The switching of the spin of the middle mode upon varying the wave vector is a consequence of the interaction of magnon excitations through the Swihart mode and anticrossing, which occurs in the areas of intersection of the bare magnon and Swihart branches. Note that such an effect was also demonstrated in \cite{Kamra2017}, but only for the case of a ferrimagnet, where the intersection of the bare [without taking into account the dipolar interaction] magnon spectrum branches is possible. In our case, anticrossing and spin switching occur for an antiferromagnet, where the magnon branches themselves do not intersect, and their interaction occurs through the Swihart mode. It is also worth mentioning that, as a result of neglecting the squeezing terms, the effect of an increase in the magnon spin above $\hbar$ at small $k$ demonstrated in \cite{Kamra2017} is not observed in our case.

In the region of strong magnon-photon mixing $k \approx k_K$, the group velocities of the magnon-polariton branches $v_j(\bm k) = d \Omega_j/d \bm k$ are $v_j \lesssim c/4$. This velocity is much larger than that of the fastest magnon group velocities reported~\cite{Chen2017}. Thus, due to the combination of non-zero spin of order of $\hbar$, high group velocities, high frequencies, and low dissipation, the hybrid antiferromagnetic magnon-polaritons, predicted in our work, should be extremely efficient for transmitting magnetic and optical signals, opening up prospects for ultrafast and energy-efficient data processing.

\section{Magnon-polaritons: Classical view}
\label{sec:classical}

\subsection{Classical derivation of the dispersion relations}
\label{sub:classical_dispersion}

Now we shall use an alternative classical approach to obtain the magnon-polariton eigenfrequencies and mode functions. The Landau-Lifshitz-Gilbert equation for the  magnetizations of the two sublattices $\bm M_{1,2}$ takes the form:
  \begin{align}
     \dfrac{\partial\bm M_i}{\partial t}=-\gamma\mu_0\bm M_i\times\bm H_i{+\dfrac{\alpha}{M_s}\bm M_i\times\dfrac{\partial\bm M_i}{\partial t} },~~i\in\{1,2\},
     \label{LLG}
 \end{align}
 in which 
  \begin{align}
     \bm H_i=\dfrac{K}{M_s}M_{iz}\bm e_z-\dfrac{J_{AF}}{M_s}\bm M_{\overline{i}}+\bm H_0+\bm H^{d}
     \label{H eff}
 \end{align}
 is the total effective magnetic field acting on the magnetization of the sublattice $i$, $\overline{i}=2(1)$ if $i=1(2)$, and $\alpha$ is the Gilbert damping constant. Substituting Eq. \eqref{h0} and a classical analogue of Eq. \eqref{H_d} into Eq. \eqref{H eff} and then linearizing Eq. (\ref{LLG}) with respect to the magnon amplitudes, we obtain 
 \begin{widetext}
  \begin{align}
     \left(\begin{array}{cccc}
          i\omega&0&-\gamma\mu_0 (F_++N_y M_s){+i\alpha\omega}&-\gamma\mu_0 (J_{AF}+N_y M_s)\\
         0 & i\omega&\gamma\mu_0 (J_{AF}+N_y M_s)&\gamma\mu_0 (F_-+N_y M_s){-i\alpha\omega}\\
         \gamma\mu_0 (F_++N_x M_s){-i\alpha\omega}&\gamma\mu_0 (J_{AF}+N_x M_s)&i\omega&0\\
         -\gamma\mu_0 (J_{AF}+N_x M_s)&-\gamma\mu_0 (F_-+ N_x M_s){+i\alpha\omega}&0&i\omega
     \end{array}\right)\left(\begin{array}{c}
        \tilde M_{1x}\\
        \tilde M_{2x}\\
           \tilde M_{1y}\\
           \tilde M_{2y}
    \end{array}\right)=0,
     \label{M_matrix}
 \end{align}
 \end{widetext}
 where $F_{\pm}=K+J_{AF}\pm H_0$. The eigenfrequencies of the collective excitations in the S/AF/S heterostructure are to be obtained from Eq.~(\ref{M_matrix}). Let us note that neglecting the demagnetization fields [that is, taking the limiting case $N_x \to 0$, $N_y\to 0$ in Eq.~(\ref{M_matrix})], the magnon eigenfrequencies would take the form
 \begin{align}
     \omega_{\pm}=\gamma\mu_0 \Bigl(\sqrt{(K+J_{AF})^2-J_{AF}^2}\pm H_0\Bigr),
     \label{omega_pm}
 \end{align}
  which is a well-known result for the magnon modes in an easy-axis antiferromagnet \cite{PhysRev.85.329,Rezende2019} and which was also obtained by our quantum approach as a limiting case of Eq. \eqref{E_k}.

  From now on, we will use the shortened notations $N_x=1$ and $N_y \equiv N$. The eigenfrequencies, calculated from Eq.~(\ref{M_matrix}), take the form
 \begin{align}\omega_{1,2}&=\gamma\mu_0\Bigg(H_0^2+\dfrac{\omega_{01}^2+\omega_{02}^2}{2\gamma^2\mu_0^2}\nonumber\\
 &\pm\sqrt{K^2M_s^2(N-1)^2+2H_0^2\dfrac{\omega_{01}^2+\omega_{02}^2}{\gamma^2\mu_0^2}}\Bigg)^{\frac{1}{2}},
     \label{eigen_general}
 \end{align}
 where 
 \begin{subequations}
 \begin{align}
    \omega_{01}&=\gamma\mu_0\sqrt{K(2J_{AF}+K+2M_s)},
     \label{eigen_01}\\
    \omega_{02}&=\gamma\mu_0\sqrt{K(2J_{AF}+K+2NM_s)},
     \label{eigen_02}
 \end{align}
 \end{subequations}
 are the eigenfrequencies in the case with no external magnetic field, $H_0=0$. In Eqs.~(\ref{eigen_01})-(\ref{eigen_02}), corrections on the order of $\alpha^2$ arising from Gilbert damping are neglected. The first-order corrections yield the decay rates of the modes $\omega_{01,2}$:
\begin{align}
\kappa_{01} &= \alpha \gamma \mu_0 (J_{AF} + K + M_s), \\
\kappa_{02} &= \alpha \gamma \mu_0 (J_{AF} + K + N M_s).
\end{align}
Using the same parameters as in Fig.~\ref{fig:eigenfrequencies_qv} and taking $\alpha \sim 10^{-4}$, the magnitude of these decay rates can be estimated as $\kappa_{01,2} \sim 1\ \mathrm{GHz}$.

This estimate enables us to evaluate the cooperativity of the magnon-polaritons, defined as $C = 4g^2/(\kappa_m \kappa_{ph})$, where $\kappa_m$ and $\kappa_{ph}$ are the magnon and photon decay rates, respectively. For our S/AF/S heterostructure, with $\kappa_{ph} \approx 3\ \mathrm{MHz}$ \cite{Huebl2013,Silaev2022} and $\kappa_m \approx \kappa_{01,2}(k=0) \approx 1\ \mathrm{GHz}$, we obtain an exceptionally high cooperativity $C \approx 10^7$. This value significantly exceeds the cooperativities reported for AF-based light--matter systems \cite{Kritzell2023,Sivarajah2019,Bialek2020,Baydin2023,Blank2023}, as well as those in S/F/S structures, where $C \sim 10^4$ has been reported \cite{Silaev2022}.

Furthermore, the parameter $U = \sqrt{C\,(g/\omega)}$, which quantifies the degree of coherence in ultrastrongly coupled systems \cite{Forn_Diaz2019}, also attains a very large value in our case, $U \sim 10^3$. This is comparable to the highest values reported for superconducting qubits. 
 
As the demagnetization factor $N$ is expressed by Eq. (\ref{N}) and depends on $\omega$ and $k$,
 Eq. (\ref{eigen_general}) and Eq. (\ref{eigen_02}) 
 are actually implicit equations for $\omega_{1,2}(k)$ and $\omega_{02}(k)$, respectively.  Figure~\ref{fig:polarization} shows 
the dispersion relations, calculated for the structure where the antiferromagnet is $\mathrm{MnF}_2$ with the same parameters addressed in Sec.~\ref{sec:quantum}. Here we again consider the bulk volume configuration, i.e., $\theta=0$. We note that according to the quantum formalism, the coupling constants between magnons and Swihart photon vanish in the Damon-Eshbach configuration $\theta=\pi/2$. The dispersion relations calculated via the quantum approach from Eq.~\eqref{Omega_qv} and via the classical approach from Eq.~\eqref{eigen_general} lead to identical pictures. In the following text of this section, we will refer only to expressions for the eigenmodes obtained by the classical approach.

\subsection{Polarization of the eigenmodes}
\label{sub:polarization}

\begin{figure} 
    \centering
    \includegraphics[width=\linewidth]{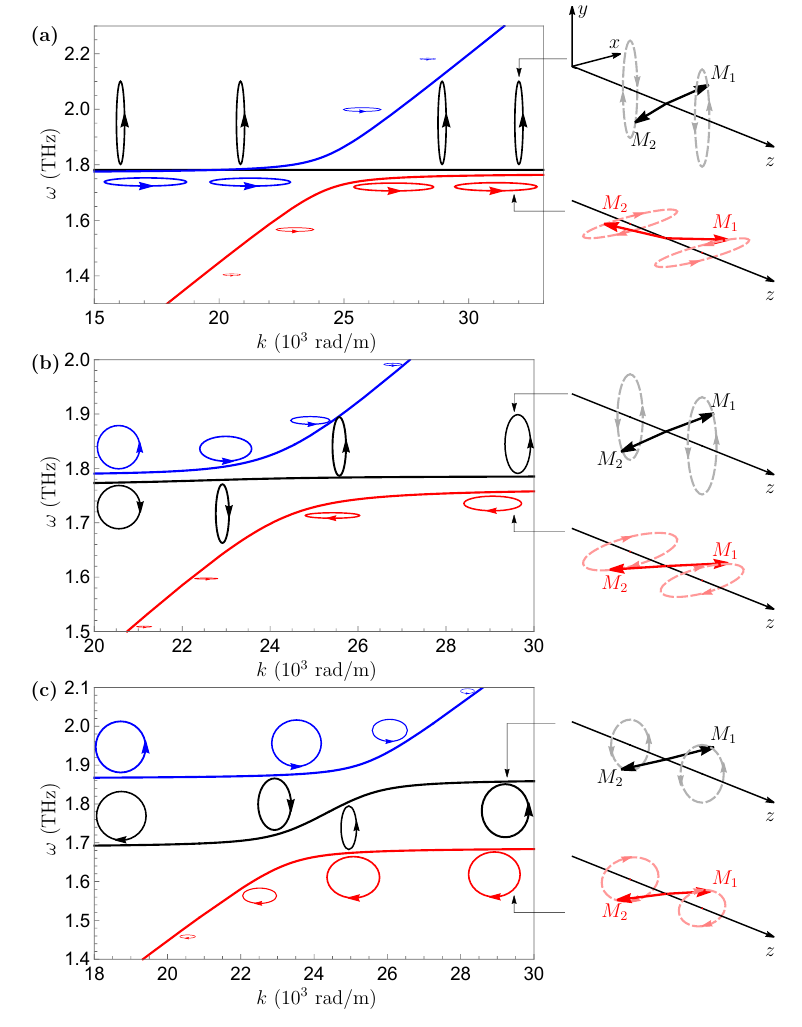}
        \caption{The magnetization configuration on the eigenmodes in the cases of zero external field [(a)], moderate external field $\mu_0H_0=0.05$ T [(b)], and strong external field $\mu_0H_0=0.5$ T [(c)].  The left side of each panel shows the ellipses of $\bm M_1$ precession at different points on the eigenmodes, plotted next to those points. The colors of the ellipses are the same as the colors of the corresponding modes. The right side shows the extended pictures, involving precession of the magnetizations at both sublattices, for some of the ellipses from the left side, marked by arrows. Parameters are the same as in Fig.~\ref{fig:eigenfrequencies_qv}.}
    \label{fig:polarization}
\end{figure}

The classical language allows for the discussion of the polarization of the eigenmodes. Its amplitudes on a particular mode are obtained as the solution $(\tilde M_{1x},
        \tilde M_{2x},
           \tilde M_{1y},
           \tilde M_{2y})^T$ of Eq.~(\ref{M_matrix}) with substituted frequency of that mode as $\omega$. 
The calculated magnetization configuration at some particular points of the eigenmodes is presented in Fig.~\ref{fig:polarization}.  
Let us first look at the left-hand side of each panel. It shows the frequency dispersion on a range that is convenient for displaying the polarizations. On this side of Fig.~\ref{fig:polarization} we show the magnetization of only one sublattice, $\bm M_1$. It is presented via ellipses plotted with the same colors as the corresponding modes and with their centers situated at  approximately the same $k$ as the points they correspond to. The horizontal and vertical semiaxes of the ellipses represent the amplitudes $\tilde M_{1x}$ and $\mathrm{Im}~ \tilde M_{1y}$, respectively. The arrows on the ellipses show the direction of the magnetization precession. The size of the ellipses was obtained from the condition of the value $\sqrt{|\tilde M_{1x}|^2+|\tilde M_{1y}|^2}$ for each ellipse being proportional to the magnonic part of the magnon-polariton at the corresponding point on the mode. That is why the ellipses become smaller at the components of the eigenmodes originating from the Swihart mode. The right-hand side of Fig.~\ref{fig:polarization} depicts the relative configuration of the magnetizations of the two sublattices at the points corresponding to the ellipses from the left-hand side, marked by arrows. Here we schematically show the ellipses for both the magnetizations $\bm M_1$ and $\bm M_2$ together with these magnetizations themselves at some random $\bm\rho$ and $t$.    

First, let us focus on the case $H_0=0$, presented in Fig.~\ref{fig:polarization}(a). The magnetization amplitudes on the dispersionless magnon mode $\omega_{01}$ (\ref{eigen_01}) (the black mode) take the form 
\begin{align}
    \tilde M_{2x}=\tilde M_{1x}, ~~\tilde M_{1,2y}=\pm i\dfrac{\omega_{01}}{\gamma\mu_0K}\tilde M_{1x},
    \label{middle_0}
\end{align}
which leads to the magnetizations on this mode having elliptical polarization aligned with the $y$ axis. As can be seen from Eq.~(\ref{middle_0}) and is illustrated in the right side of panel (a), on this mode $\bm M_1$ and $\bm M_2$ precess in opposite directions. The net magnetization is oriented along the $x$ axis and has a zero $y$-component, $\tilde M_y=\tilde M_{1y}+\tilde M_{2y}=0$. Since the only nonzero magnetic field components of the Swihart mode are $H_{Sw,y,z}$, it means that it cannot be coupled to the corresponding magnon mode, in full agreement with the quantum approach.  The components of magnetizations on hybrid magnon-polariton modes, plotted as the red and blue curves, obey the following relations:
\begin{widetext}
\begin{align}
    &\tilde M_{2x}=-\tilde M_{1x}, \nonumber\\
    &\tilde M_{1y}=\tilde M_{2y}=i\gamma\mu_0K\sqrt{\dfrac{2\mu_0\varepsilon_{AF}(d_{AF}+\lambda_{\rm{eff}})}{A_1+d_{AF}k^2\pm\sqrt{(A_1-d_{AF}k^2)^2+4\mu_0\varepsilon_{AF}d_{AF}^2k^2\gamma^2\mu_0^2KM_s\cos{2\theta}}}}\tilde M_{1x},
\nonumber\\
&A_1=\mu_0\varepsilon_{AF}(\omega_{01}^2(d_{AF}+\lambda_{\rm{eff}})-2\lambda_{\rm{eff}}\gamma^2\mu_0^2KM_s).
    \label{middle_12}
\end{align}
\end{widetext}
There, the sign $+(-)$ in the expression for $\tilde M_{1,2y}$ corresponds to the blue (red) mode.
Thus, the net magnetization on those modes is aligned with the $y$ axis [see the bottom part of the right side of panel (a)], giving way to magnon-photon coupling. On the red and blue modes, the precession directions of $\bm M_1$ and $\bm M_2$ are also opposite to each other, and in the displayed range of $k$, both magnetizations are polarized elliptically along the $x$ axis. However, the magnetizations on the red mode have linear along $y$ polarizations at $k\to 0$, which are then gradually transformed to elliptical along $x$ going through stage of circular polarizations at $k\sim 2000$ rad/m, and the magnetizations on the blue mode become polarized linearly along $x$ at $k\to\infty$, as it should be for the spectra regions originating from the photon mode. Let us note that in the case $H_0=0$, the precession direction of each sublattice magnetization remains constant for all the wave vector values on all the eigenmodes.

Now we turn to the case of non-zero external magnetic field, see Figs.~\ref{fig:polarization}(b)-(c). In this case, the magnetization amplitudes take the form:
\begin{align}
    &\tilde M_{2x}=\dfrac{\tilde M_{1x}}{A_2}\nonumber\\
    &\times\Bigl((F_++M_s)(\gamma^2\mu_0^2H_0^2-\omega_{02}^2)+(M_sN+F_-)\omega^2\Bigr),\nonumber\\
    &\tilde M_{1y}=\dfrac{i\omega\gamma\mu_0}{A_2}\tilde M_{1x}\nonumber \\
    &\times \Bigl(KM_s(N-1)+H_0(2J_{AF}+M_s(N+1))\Bigr),\nonumber\\
    &\tilde M_{2y}=\dfrac{i\omega}{A_2}\tilde M_{1x}\nonumber\\
    &\times \Bigl(\gamma^2\mu_0^2(H_0+K)(H_0+2J_{AF}+K+M_s(N+1))+\omega^2\Bigr),\nonumber\\
    &A_2=(J_{AF}+M_s)(\omega_{02}^2-\gamma^2\mu_0^2H_0^2)-(J_{AF}+NM_s)\omega^2.
    \label{M_ampl}
\end{align}
To obtain the magnetizations on a particular eigenmode, we have to substitute the eigenfrequencies from Eq.~(\ref{eigen_general}). In the general case the explicit expressions for the magnetizations amplitudes are too complicated, but in the limits $k=0$ and $k \to\infty$, Eq.~(\ref{M_ampl}) gives us these explicit expressions, as $N(k=0)=d_{AF}/(d_{AF}+\lambda_{\rm{eff}})$ and $N(k \to\infty)=d_{AF}\sin^2 \theta/(d_{AF}+\lambda_{\rm{eff}})$.
Let us discuss the magnetizations at a moderate external field [panel (b)] and then analyze what changes with increasing field. In the middle black mode, the magnetization $\bm M_1$ is polarized elliptically along $x$ at small $k$ (not shown in the figure), then as the wave vector increases, the polarization gets circular and then elliptical along $y$. Approaching the anticrossing point, the polarization becomes linear because the mixing of two nearly circularly polarized magnonic modes in the region of coupling with the Swihart mode becomes much stronger.  At this moment, there is no precession direction, which is why, when the polarization becomes elliptical along $y$ again after the anticrossing region, the precession direction can switch to the opposite one, as shown in the figure. On the red mode before the anticrossing, the configuration of $\bm M_1$ is generally the same as at $H_0=0$, but moving through the anticrossing point is accompanied by the same process of the polarization becoming linear and switching of the precession direction, as on the black mode. On the blue mode for $\bm M_1$, we see only transformation from elliptical along $y$ to circular and elliptical along $x$ polarization without switching of the precession direction. Still, for the magnetization of the other sublattice $\bm M_2$, such switching also occurs. 

The right side of Fig.~\ref{fig:polarization}(b) illustrates that after the anticrossing $\bm M_1$ and $\bm M_2$ precess in the same direction and are oriented nearly oppositely on each of the black and red modes. However, in the case of non-zero $H_0$, the absolute values of the magnetizations $M_1$ and $M_2$ are not equal to each other, which is shown in the figure by different sizes of the ellipses for the two sublattices. Thus, now the net magnetization on each of the modes has a non-zero $y$-component, and both magnon modes are coupled with the Swihart mode (although it might not be clear from the black curve on the left side of panel (b), this mode also depends on $k$ and represents another magnon-polariton branch). 

Now let us analyze the case of a strong external field, Fig.~\ref{fig:polarization}(c). On the black mode the differences from the case of moderate field are that (i) the abrupt change of the magnetization configuration takes place in a narrower range of $k$ around the anticrossing point; (ii) the polarizations far before and after the anticrossing are closer to circular. The polarizations also become closer to circular on the before-anticrossing and after-anticrossing parts of blue and red modes, respectively (i.e., on the magnon-originated parts of these branches). The switching of the precession direction of $\bm M_1$ on the red mode and $\bm M_2$ on the blue mode now takes place farther from the anticrossing (outside of the area displayed in the figure) on the photon-originated parts of these modes. The right sides of panels (b) and (c) are topologically the same; the only difference is in the shape of the ellipses.

\section{Conclusions}
\label{sec:conclusions}
In this work, we have proposed and theoretically investigated a platform for achieving ultrastrong magnon-photon coupling at terahertz frequencies using a S/AF/S heterostructure. Our analysis, conducted from both quantum and classical perspectives, reveals several key findings:

(i) {\it Enhanced coupling via superconductivity}. The presence of superconducting layers drastically modifies the electromagnetic environment of the antiferromagnet. The Meissner supercurrents screen the dipolar fields, leading to a significant renormalization of the magnon spectrum and enabling a strong electromagnetic interaction between the antiferromagnetic magnons and the photonic Swihart mode of the structure. The coupling strength is shown to be substantially larger than that reported for antiferromagnets in conventional terahertz cavities, placing the system at the boundary of the ultrastrong coupling regime ($g/\omega \sim 0.1$).

(ii) {\it Magnetic field control of coupling selectivity.} The coupling between magnons and photons is highly tunable by an external magnetic field. At zero field, the coupling is selective: only one of the two antiferromagnetic resonance modes couples to the photon, forming a bright magnon-polariton, while the other remains dark. Under an applied magnetic field, this selectivity is lifted, and both magnon modes couple to the Swihart mode, with the coupling becoming equal for both modes in the limit of a strong field.

In general case, the magnon-photon coupling results in three magnon-polariton branches with modified dispersions. These hybrid quasiparticles exhibit remarkable properties. First of all, the magnon-polaritons carry a non-integer average spin $\langle S_z \rangle < \hbar$, which is tunable by both the external magnetic field and the wavevector. A notable prediction is the switching of the spin sign for the middle polariton branch upon varying the wavevector, a consequence of anticrossing mediated by the Swihart mode. The second important property is the high group velocity of the magnon-polaritons. In the strong-mixing region, the group velocities of the magnon-polaritons can reach a significant fraction of the speed of light,  vastly exceeding typical magnon group velocities, which promises strong tunability of magnon transport in antiferromagnets by superconductors.
    
The classical approach based on the Landau-Lifshitz-Gilbert equation and Maxwell's equations yields results fully consistent with the quantum formalism. The classical analysis provides an intuitive picture of the magnetization precession polarization for the hybrid modes, illustrating the elliptical precession and the relative configurations of the two antiferromagnetic sublattices that underpin the coupling mechanism and the dark/bright mode behavior. 

Ultimately, the S/AF/S heterostructure emerges as a versatile platform for THz quantum magnonics. The synergy of ultrastrong coupling, magnetic tunability, and fast, low-loss magnon-polaritons opens a path to creating ultrafast, energy-efficient devices for spintronics and quantum information processing.

\begin{acknowledgments}
V.M.G., G.A.B. and I.V.B. acknowledge the support from Theoretical Physics and Mathematics Advancement Foundation “BASIS” via the project No. 23-1-1-51-1. The quantum calculation of the magnon-polariton dispersions was supported by the Russian Science Foundation via the RSF project No. 23-72-30004. The classical calculation has been obtained under the support by Grant from the ministry of science and higher education of the Russian Federation No. 075-15-2025-010. Y.M.L., X.Y.Y., and T.Y. are financially supported by the National Key Research and Development Program of China under Grant No.~2023YFA1406600 and the National Natural Science Foundation of China under Grant No.~12374109.
\end{acknowledgments}

\begin{widetext}
 \section*{Appendix A: derivation of the demagnetization tensor}

 Here we present the detailed derivation of the expression for the demagnetization tensor $\hat N$ [Eq.~(\ref{N})].  The continuity of $E_y$ and $E_z$, obtained from Eq.~(\ref{Maxwell_solutions}), at the interfaces $x=\pm d_{AF}$ gives us the following two series of conditions, respectively:
\begin{align}
    S_{1y}&=A_{y}^+e^{iAd_{AF}}+A_{y}^-e^{-iAd_{AF}}+\dfrac{\omega\mu_0}{A^2}k_z\tilde M_{x},\nonumber\\
    S_{2y}&=A_{y}^+e^{-iAd_{AF}}+A_{y}^-e^{iAd_{AF}}+\dfrac{\omega\mu_0}{A^2}k_z\tilde M_{x},
    \label{boundary_Ey}
\end{align}
and
\begin{align}
    S_{1z}&=A_{z}^+e^{iAd_{AF}}+A_{z}^-e^{-iAd_{AF}}-\dfrac{\omega\mu_0}{A^2}k_y\tilde M_{x},\nonumber\\
    S_{2z}&=A_{z}^+e^{-iAd_{AF}}+A_{z}^-e^{iAd_{AF}}-\dfrac{\omega\mu_0}{A^2}k_y\tilde M_{x}.
    \label{boundary_Ez}
\end{align}
The continuity of the components $H_z$ and $H_y$, which are obtained from  Eq. (\ref{H_final}) with the use of Eq. (\ref{Maxwell_final}), leads to
\begin{align}
    -\lambda_{\rm{eff}}^{-1}S_{1y}&=\dfrac{1}{A^2}\left(iA\alpha(A_{y}^+e^{iAd_{AF}}-A_{y}^-e^{-iAd_{AF}})+iA\dfrac{K_1}{2}(A_{z}^+e^{iAd_{AF}}-A_{z}^-e^{-iAd_{AF}})+i\omega\mu_0\dfrac{K_1}{2}\tilde M_{y}\right),\nonumber\\
    \lambda_{\rm{eff}}^{-1}S_{2y}&=\dfrac{1}{A^2}\left(iA\alpha(A_{y}^+e^{-iAd_{AF}}-A_{y}^-e^{iAd_{AF}})+iA\dfrac{K_1}{2}(A_{z}^+e^{-iAd_{AF}}-A_{z}^-e^{iAd_{AF}})+i\omega\mu_0\dfrac{K_1}{2}\tilde M_{y}\right),
     \label{boundary_Hz}
\end{align}
and
\begin{align}
    -\lambda_{\rm{eff}}^{-1}S_{1z}&=\dfrac{1}{A^2}\left(iA\beta(A_{z}^+e^{iAd_{AF}}-A_{z}^-e^{-iAd_{AF}})+iA\dfrac{K_1}{2}(A_{y}^+e^{iAd_{AF}}-A_{y}^-e^{-iAd_{AF}})+i\omega\mu_0\beta\tilde M_{y}\right),\nonumber\\
    \lambda_{\rm{eff}}^{-1}S_{2z}&=\dfrac{1}{A^2}\left(iA\beta(A_{z}^+e^{-iAd_{AF}}-A_{z}^-e^{iAd_{AF}})+iA\dfrac{K_1}{2}(A_{y}^+e^{-iAd_{AF}}-A_{y}^-e^{iAd_{AF}})+i\omega\mu_0\beta\tilde M_{y}\right).
     \label{boundary_Hy}
\end{align}
Getting rid of the amplitudes $S_{1(2)y,z}$, we can rewrite Eqs. (\ref{boundary_Ey}-\ref{boundary_Hy}) in the following matrix form:
\begin{align}
    \hat X\hat E=\hat M_E,~~\hat X=\left(\begin{array}{cccc}
        x_{1y}^+ &  x_{1y}^-& x_{1z}^+& x_{1z}^- \\
          x_{2y}^+&  x_{2y}^-&  x_{2z}^+& x_{2z}^-\\
         x_{3y}^+ &  x_{3y}^-& x_{3z}^+& x_{3z}^-\\
          x_{4y}^+ &  x_{4y}^-& x_{4z}^+& x_{4z}^-
    \end{array}\right),~~\hat E=\left(\begin{array}{c}
        A_{y}^+\\
           A_{y}^-\\
         A_{z}^+\\
           A_{z}^-
    \end{array}\right),~~\hat M_E=\dfrac{\omega\mu_0}{A^2}\left(\begin{array}{c}
        \dfrac{k\cos\theta}{\lambda_{\rm{eff}}}\tilde M_x+\dfrac{K_1}{2}i\tilde M_y\\
           -\dfrac{k\cos\theta}{\lambda_{\rm{eff}}}\tilde M_x+\dfrac{K_1}{2}i\tilde M_y\\
         \dfrac{k\sin\theta}{\lambda_{\rm{eff}}}\tilde M_x-\beta i\tilde M_y\\
           -\dfrac{k\sin\theta}{\lambda_{\rm{eff}}}\tilde M_x-\beta i\tilde M_y
    \end{array}\right).
    \label{A_matrix}
\end{align}
The coefficients in the matrix $\hat X$ in Eq. (\ref{A_matrix}) are
\begin{align}
    x_{1y}^{\pm}&=-e^{\pm iAd_{AF}}\left( \dfrac{1}{\lambda_{\rm{eff}}}\pm\dfrac{i\alpha}{A}\right),~~
    x_{1z}^{\pm}=\mp e^{\pm iAd_{AF}}\dfrac{iK_1}{2A},\nonumber\\
    x_{2y}^{\pm}&=e^{\mp iAd_{AF}}\left( \dfrac{1}{\lambda_{\rm{eff}}}\mp\dfrac{i\alpha}{A}\right),~~~~
    x_{2z}^{\pm}=\mp e^{\mp iAd_{AF}}\dfrac{iK_1}{2A},\nonumber\\
    x_{3y}^{\pm}&=\pm e^{\pm iAd_{AF}}\dfrac{iK_1}{2A},~~~~~~~~~~~~~~
    x_{3z}^{\pm}=e^{\pm iAd_{AF}}\left( \dfrac{1}{\lambda_{\rm{eff}}}\pm\dfrac{i\beta}{A}\right),\nonumber\\
    x_{4y}^{\pm}&=\pm e^{\mp iAd_{AF}}\dfrac{iK_1}{2A},~~~~~~~~~~~~~~
    x_{4z}^{\pm}=-e^{\mp iAd_{AF}}\left( \dfrac{1}{\lambda_{\rm{eff}}}\mp\dfrac{i\beta}{A}\right).
    \label{X}
\end{align}

 The components of the magnetic field amplitudes $\tilde H_{x,y}$ in the antiferromagnet can be expressed via the components of $\hat E$ and the net magnetization amplitudes using Eq.~(\ref{H_final}) and Eq.~(\ref{Maxwell_solutions}):
\begin{align}
    \tilde H_{x}&=\dfrac{k}{\omega\mu_0}\left [\sin\theta\left (A_{z}^+e^{iAx}+A_{z}^-e^{-iAx}-\dfrac{\omega\mu_0}{A^2}k\sin\theta\tilde M_x\right )-\cos\theta\left (A_{y}^+e^{iAx}+A_{y}^-e^{-iAx}+\dfrac{\omega\mu_0}{A^2}k\cos\theta\tilde M_x\right )\right ]-\tilde M_{x},\nonumber\\
    \tilde H_{y}&=-\dfrac{1}{A\omega\mu_0}\left(\beta(A_{z}^+e^{iAx}-A_{z}^-e^{-iAx})+\dfrac{K_1}{2}(A_{y}^+e^{iAx}-A_{y}^-e^{-iAx})\right)-\dfrac{\beta}{A^2}\tilde M_{y}.
    \label{H_AF}
\end{align}
As for the parameters under consideration $Ad_{AF}\ll 1$, we can average Eq.~(\ref{H_AF}) over the thickness of the AF layer, which leads to
\begin{align}
    \tilde H_{x}&=\dfrac{k}{\omega\mu_0}\left [\sin\theta\left (A_{z}^++A_{z}^-\right )-\cos\theta\left (A_{y}^++A_{y}^-\right )\right ]-\dfrac{k_{AF}^2}{A^2}\tilde M_{x},\nonumber\\
    \tilde H_{y}&=-\dfrac{1}{A\omega\mu_0}\left(\beta(A_{z}^+-A_{z}^-)+\dfrac{K_1}{2}(A_{y}^+-A_{y}^-)\right)-\dfrac{\beta}{A^2}\tilde M_{y},
    \label{H_AF_final}
\end{align}
or, in the matrix form,
\begin{align}
    \hat H&=\hat \Gamma\hat E+\hat\Gamma_M\hat {M},
    ~~\hat H=\left(\begin{array}{c}
        \tilde H_{x}\\
           \tilde H_{y}
    \end{array}\right),~~\hat M=\left(\begin{array}{c}
        \tilde M_{x}\\
           \tilde M_{y}
    \end{array}\right),\nonumber\\
    \hat\Gamma&=\dfrac{1}{\omega\mu_0}\left(\begin{array}{cccc}
        -k\cos\theta & -k\cos\theta & k\sin\theta & k\sin\theta\\
        -\dfrac{K_1}{2A}& \dfrac{K_1}{2A}&-\dfrac{\beta}{A}&\dfrac{\beta}{A}
    \end{array}\right),~~\hat\Gamma_M=\left(\begin{array}{cc}
        -\dfrac{k_{AF}^2}{A^2} & 0 \\
         0&-\dfrac{\beta}{A^2}
    \end{array}\right).
    \label{matrix H}
\end{align}
Then in Eq.~(\ref{matrix H}) we can substitute $\hat E$ expressed via $\hat {M}$ from  Eq. (\ref{A_matrix}):
\begin{align}
    \hat E=\hat X^{-1}\hat M_E=\hat X^{-1}\hat M_t\hat {M}, ~~\hat M_t=\dfrac{\omega\mu_0}{A^2}\left(\begin{array}{cc}
        \dfrac{k\cos\theta}{\lambda_{\rm{eff}}}&i\dfrac{K_1}{2} \\
         -\dfrac{k\cos\theta}{\lambda_{\rm{eff}}}&i\dfrac{K_1}{2}\\
         \dfrac{k\sin\theta}{\lambda_{\rm{eff}}}&-i\beta\\        
         -\dfrac{k\sin\theta}{\lambda_{\rm{eff}}}&-i\beta
    \end{array} \right),
    \label{E_M}
\end{align}
which gives us the following relation between $\hat H$ and $\hat M$:
\begin{align}
    \hat H=(\hat\Gamma\hat X^{-1}\hat M_t+\hat\Gamma_M)\hat M.
    \label{H_M2}
\end{align}
Comparing Eq. (\ref{H_M2}) with Eq. (\ref{H_M}), we obtain 
\begin{align}
    \hat N=-(\hat\Gamma\hat X^{-1}\hat M_t+\hat\Gamma_M),
\end{align}
which leads to the final expression for the demagnetization tensor (up to the zero order with respect to $Ad_{AF}$ and $ k\lambda_{\rm{eff}}$):
\begin{align}
     \hat N=\left(\begin{array}{cc}
       N_x  & 0 \\
         0 & N_y
     \end{array}\right),~~~
     N_x=1,~~~
     N_y=\dfrac{d_{AF}\left(d_{AF}k^2\sin^2\theta-k_{AF}^2(d_{AF}+\lambda_{\rm{eff}})\right)}{(d_{AF}+\lambda_{\rm{eff}})\left(d_{AF}k^2-k_{AF}^2(d_{AF}+\lambda_{\rm{eff}})\right)}.
     \label{N1}
 \end{align}

\section*{Appendix B: magnon-polariton eigenfrequencies from the quantum approach}

The magnon-polariton eigenfrequencies are obtained from the following equation:
\begin{align}
    &\omega^6-(\tilde{\omega}_1^2+\tilde{\omega}_2^2+\Omega_s^2)\omega^4+(\tilde{\omega}_1^2\tilde{\omega}_2^2+\tilde{\omega}_1^2\Omega_s^2+\tilde{\omega}_2^2\Omega_s^2-4|g_1|^2\tilde{\omega}_1\Omega_s-4|g_2|^2\tilde{\omega}_2\Omega_s)\omega^2\nonumber\\
    &+\tilde{\omega}_1\tilde{\omega}_2\Omega_s(4|g_1|^2\tilde{\omega}_2+4|g_2|^2\tilde{\omega}_1-\tilde{\omega}_1\tilde{\omega}_2\Omega_s)=0,
\end{align}
the solutions of which are
\begin{align}
    \Omega_1&=\frac{1}{\sqrt{3}}\sqrt{\tilde{\omega}_1^2+\tilde{\omega}_2^2+\Omega_s^2-\frac{\sqrt[3]{2}\tilde{C}}{\tilde{A}}+\frac{\tilde{A}}{\sqrt[3]{2}}},\nonumber\\
    \Omega_2&=\frac{1}{\sqrt{3}}\sqrt{\tilde{\omega}_1^2+\tilde{\omega}_2^2+\Omega_s^2+\frac{(1+i\sqrt{3})\tilde{C}}{2^{\frac{2}{3}}\tilde{A}}-\frac{(1-i\sqrt{3})\tilde{A}}{2\sqrt[3]{2}}},\nonumber\\
    \Omega_3&=\frac{1}{\sqrt{3}}\sqrt{\tilde{\omega}_1^2+\tilde{\omega}_2^2+\Omega_s^2+\frac{(1-i\sqrt{3})\tilde{C}}{2^{\frac{2}{3}}\tilde{A}}-\frac{(1+i\sqrt{3})\tilde{A}}{2\sqrt[3]{2}}},
    \label{Omega_qv}
    \end{align}
    where
    \begin{align}
    \tilde{A}&=\sqrt[3]{\tilde{B}+\sqrt{4\tilde{C}^3+\tilde{B}^2}},\nonumber\\
    \tilde{B}&=2(\tilde{\omega}_1^6+\tilde{\omega}_2^6+\Omega_s^6)-3(\tilde{\omega}_1^4\tilde{\omega}_2^2+\tilde{\omega}_1^4\Omega_s^2+\tilde{\omega}_1^2\tilde{\omega}_2^4+\tilde{\omega}_1^2\Omega_s^4+\tilde{\omega}_2^4\Omega_s^2+\tilde{\omega}_2^2\Omega_s^4)+12\tilde{\omega}_1^2\tilde{\omega}_2^2\Omega_s^2\nonumber\\
    &+36(|g_1|^2\tilde{\omega}_1\Omega_s(\tilde{\omega}_1^2+\Omega_s^2)+|g_2|^2\tilde{\omega}_2\Omega_s(\tilde{\omega}_2^2+\Omega_s^2))-72\tilde{\omega}_1\tilde{\omega}_2\Omega_s(|g_1|^2\tilde{\omega}_2+|g_2|^2\tilde{\omega}_1),\nonumber\\
    \tilde{C}&=\tilde{\omega}_1^2\tilde{\omega}_2^2+\tilde{\omega}_1^2\Omega_s^2+\tilde{\omega}_2^2\Omega_s^2-\tilde{\omega}_1^4-\tilde{\omega}_2^4-\Omega_s^4-12\Omega_s(|g_1|^2\tilde{\omega}_1+|g_2|^2\tilde{\omega}_2).
\end{align}

\section*{Appendix C: calculation of the average spin on the  eigenmodes}

In this Appendix for simplicity we will use the notation $\hbar=1$. In terms of the bosonic operators $\hat a_i(\bm r)$ the average magnon spin $\langle S_z \rangle$ takes the form [see Eq. (\ref{HP2})]
\begin{align}
    \langle S_z \rangle=\langle S_{1z}\rangle+\langle S_{2z}\rangle=\frac{1}{2d_{AF}}\int d^2{\bm \rho} dx\sum_{{\bm k}{\bm k}'}\Bigl(\langle\hat{a}^{\dagger}_{1,{\bm k}}\hat{a}_{1,{\bm k}'}\rangle e^{i({\bm k}'-{\bm k})\cdot{\bm \rho}}-\langle\hat{a}^{\dagger}_{2,{\bm k}}\hat{a}_{2,{\bm k}'}\rangle e^{i({\bm k}'-{\bm k})\cdot{\bm \rho}}\Bigr)=\sum_{{\bm k}}\Bigl(\langle\hat{a}^{\dagger}_{1,{\bm k}}\hat{a}_{1,{\bm k}}\rangle-\langle\hat{a}^{\dagger}_{2,{\bm k}}\hat{a}_{2,{\bm k}}\rangle\Bigr),
\end{align}
which can be rewritten via the magnon operators $\hat{m}_{1(2),{\bm k}}$ with the use of Eq. (\ref{a-m}):
\begin{align}
    \langle S_z \rangle=\sum_{{\bm k}}\Bigl(\langle\hat{m}^{\dagger}_{1,{\bm k}}\hat{m}_{1,{\bm k}}\rangle-\langle\hat{m}^{\dagger}_{2,{\bm k}}\hat{m}_{2,{\bm k}}\rangle\Bigr).
\end{align}
The photonic part of the excitation does not contribute to $\langle S_z \rangle$ because it is linearly polarized in the plane perpendicular to $\bm k$. The operators $\hat{m}_{1(2),{\bm k}}$ can be expressed via the new magnon operators $\tilde{m}_{1(2),{\bm k}}$ (\ref{m_new}) as
\begin{align}
\hat{m}_{1,{\bm k}}=\frac{h_-}{\sqrt{h_-^2+\alpha_-^2}}\tilde{m}_{1,{\bm k}}+\frac{\alpha_-}{\sqrt{h_-^2+\alpha_-^2}}\tilde{m}_{2,{\bm k}},~~~~
\hat{m}_{2,{\bm k}}=\frac{h_-}{\sqrt{h_-^2+\alpha_+^2}}\tilde{m}_{1,{\bm k}}+\frac{\alpha_+}{\sqrt{h_-^2+\alpha_+^2}}\tilde{m}_{2,{\bm k}},
\label{m}
\end{align}
where $\alpha_{\pm}=\mu_0\gamma H_0\pm\sqrt{(\mu_0\gamma H_0)^2+h_-^2}$. Then for the average spin, we obtain
\begin{align}
    \langle S_z \rangle&=\sum_{{\bm k}}\Bigl[\Bigl(\frac{h_-^2}{h_-^2+\alpha_-^2}-\frac{h_-^2}{h_-^2+\alpha_+^2}\Bigr)\langle\tilde{m}^{\dagger}_{1,{\bm k}}\tilde{m}_{1,{\bm k}}\rangle+\Bigl(\frac{\alpha_-^2}{h_-^2+\alpha_-^2}-\frac{\alpha_+^2}{h_-^2+\alpha_+^2}\Bigr)\langle\tilde{m}^{\dagger}_{2,{\bm k}}\tilde{m}_{2,{\bm k}}\rangle\nonumber\\
    &+\Bigl(\frac{h_-\alpha_-}{h_-^2+\alpha_-^2}-\frac{h_-\alpha_+}{h_-^2+\alpha_+^2}\Bigr)\langle\tilde{m}^{\dagger}_{1,{\bm k}}\tilde{m}_{2,{\bm k}}+\mathrm{H. c.}\rangle\Bigr].
    \label{S_m}
\end{align}

As in the case $k\neq 0$ the magnons are coupled to the Swihart mode and the operators $\tilde{m}_{1(2),{\bm k}}$ do not correspond to the eigenstates of the Hamiltonian (\ref{H_tot}), we need to rewrite Eq. (\ref{S_m}) in terms of the operators which diagonalize $\hat{H}_{ tot}$. For this purpose we express the Hamiltonian in the form
\begin{align}
    \hat{H}_{ tot}=\sum_{k>0}\hat{\psi}^{\dagger}_{ k}{\cal H}_m\hat{\psi}_{k},
\end{align}
where
\begin{align}
    {\cal H}_m=\begin{pmatrix}
        \tilde{\omega}_1&0&0&0&g^{\ast}_1&g^{\ast}_1\\  0& \tilde{\omega}_2&0&0&g^{\ast}_2&g^{\ast}_2\\
        0&0&\tilde{\omega}_1&0&g_1&g_1\\
        0&0&0&\tilde{\omega}_2&g_2&g_2\\
        g_1&g_2&g^{\ast}_1&g^{\ast}_2&\Omega_s&0\\
        g_1&g_2&g^{\ast}_1&g^{\ast}_2&0&\Omega_s
    \end{pmatrix},~~~~~\hat{\psi}_{ k}=\begin{pmatrix}
        \tilde{m}_{1,{ k}}\\\tilde{m}_{2,{ k}}\\\tilde{m}^{\dagger}_{1,-{ k}}\\\tilde{m}^{\dagger}_{2,-{ k}}\\\hat p_{k}\\\hat p^{\dagger}_{- k}
    \end{pmatrix}.
\end{align}
We then diagonalize $\hat{H}_{ tot}$ with the Bogoliubov transformation
\begin{align}
   \hat{\psi}_{k}=\hat S \hat b_{ k},
   \label{psi_beta}
\end{align}
where $\hat b_{k}=(\beta_{1, k}, \beta_{2, k}, \beta_{3, k}, \beta^{\dagger}_{1,- k}, \beta^{\dagger}_{2,- k}, \beta^{\dagger}_{3,- k})^T$ is the vector of magnon-polariton operators $\beta_{i, k}$ satisfying the relation
\begin{align}
    [\hat b_{ k},\hat{H}_{ tot}]=\hat \Omega\hat b_{k},~~~~\hat \Omega=\begin{pmatrix}
        \Omega_1&0&0&0&0&0\\0&\Omega_2&0&0&0&0\\0&0&\Omega_3&0&0&0\\0&0&0&-\Omega_1&0&0\\0&0&0&0&-\Omega_2&0\\0&0&0&0&0&-\Omega_3
    \end{pmatrix},
    \label{Omega}
\end{align}
$\Omega_{1,2,3}$ are the magnon-polariton eigenfrequencies. Let us define the matrix $\hat\xi=\hat S^{-1}$. The condition expressed by Eq. (\ref{Omega}) gives us that the rows of $\hat\xi$ are the eigenvectors of the matrix $\hat Y {\cal H}^T_m$, $\hat Y=\mathrm{diag}(1,1,-1,-1,1,-1)$, and $\pm\Omega_{1,2,3}$ are the eigenvalues of this matrix. The commutation relations for the magnon-polariton operators read
\begin{align}
    [\beta_{i,k}, \beta_{j,k'}^{\dagger}]=\hat Y_{2ii}\delta_{ij}\delta_{kk'},~~~~\hat Y_2=\mathrm{diag}(1,1,1,-1,-1,-1),
\end{align}
which leads to 
\begin{align}
    \hat S=\hat Y\hat \xi^{\dagger}\hat Y_2.
    \label{S}
\end{align}

Then let us rewrite Eq. (\ref{S_m}) in terms of the components of the vector $\hat{\psi}_{ k}$:
\begin{align}
    \langle S_z \rangle&=\sum_{{ k>0}}\Bigl[\Bigl(\frac{h_-^2}{h_-^2+\alpha_-^2}-\frac{h_-^2}{h_-^2+\alpha_+^2}\Bigr)\langle\psi_{1k}^{\dagger}\psi_{1 k}+\psi_{3k}^{\dagger}\psi_{3 k}-1\rangle+\Bigl(\frac{\alpha_-^2}{h_-^2+\alpha_-^2}-\frac{\alpha_+^2}{h_-^2+\alpha_+^2}\Bigr)\langle\psi_{2k}^{\dagger}\psi_{2k}+\psi_{4 k}^{\dagger}\psi_{4 k}-1\rangle\nonumber\\
    &+\Bigl(\frac{h_-\alpha_-}{h_-^2+\alpha_-^2}-\frac{h_-\alpha_+}{h_-^2+\alpha_+^2}\Bigr)\langle\psi_{1k}^{\dagger}\psi_{2 k}+\psi_{2k}^{\dagger}\psi_{1 k}+\psi_{4k}^{\dagger}\psi_{3 k}+\psi_{3k}^{\dagger}\psi_{4 k}\rangle\Bigr]\nonumber\\
    &=\sum_{{ k>0}}\Bigl[\Bigl(\frac{h_-^2}{h_-^2+\alpha_-^2}-\frac{h_-^2}{h_-^2+\alpha_+^2}\Bigr)\langle\psi_{1k}^{\dagger}\psi_{1k}+\psi_{3 k}^{\dagger}\psi_{3 k}\rangle+\Bigl(\frac{\alpha_-^2}{h_-^2+\alpha_-^2}-\frac{\alpha_+^2}{h_-^2+\alpha_+^2}\Bigr)\langle\psi_{2k}^{\dagger}\psi_{2k}+\psi_{4k}^{\dagger}\psi_{4k}\rangle\nonumber\\
    &+\Bigl(\frac{h_-\alpha_-}{h_-^2+\alpha_-^2}-\frac{h_-\alpha_+}{h_-^2+\alpha_+^2}\Bigr)\langle\psi_{1k}^{\dagger}\psi_{2k}+\psi_{2 k}^{\dagger}\psi_{1 k}+\psi_{4k}^{\dagger}\psi_{3 k}+\psi_{3k}^{\dagger}\psi_{4 k}\rangle\Bigr].
    \label{S_psi}
\end{align}
The final step to obtain the expressions for the average spin on the eigenmodes is to rewrite Eq. (\ref{S_psi}) via the operators $\beta_{i,k}$. From Eq. (\ref{psi_beta}) we get
\begin{align}
   \psi_{n k}=\sum_{i=1}^6 S_{ni}b_{i k},
\end{align}
then Eq. (\ref{S_psi}) takes the form
\begin{align}
    \langle S_z \rangle&=\sum_{{ k>0}}\sum_{i,j=1}^6 \Bigl[\Bigl(\frac{h_-^2}{h_-^2+\alpha_-^2}-\frac{h_-^2}{h_-^2+\alpha_+^2}\Bigr)(S_{1j}^{\ast}S_{1i}+S_{3j}^{\ast}S_{3i})+\Bigl(\frac{\alpha_-^2}{h_-^2+\alpha_-^2}-\frac{\alpha_+^2}{h_-^2+\alpha_+^2}\Bigr)(S_{2j}^{\ast}S_{2i}+S_{4j}^{\ast}S_{4i})\nonumber\\
    &+\Bigl(\frac{h_-\alpha_-}{h_-^2+\alpha_-^2}-\frac{h_-\alpha_+}{h_-^2+\alpha_+^2}\Bigr)(S_{1j}^{\ast}S_{2i}+S_{2j}^{\ast}S_{1i}+S_{4j}^{\ast}S_{3i}+S_{3j}^{\ast}S_{4i})\Bigr]\langle b_{j k}^{\dagger}b_{i k}\rangle.
    \label{S_b}
\end{align}
If we consider a state of the system with one magnon-polariton in the mode $\Omega_i(k)$, the only non-zero contribution to the sum in Eq. (\ref{S_b}) will be from $\langle b_{i k}^{\dagger}b_{i k}\rangle=1$. Then the final expression for the average spin on the eigenmode $\Omega_i(k)$ reads
\begin{align}
    \langle S_z \rangle_{i,k}&=\Bigl[\Bigl(\frac{h_-^2}{h_-^2+\alpha_-^2}-\frac{h_-^2}{h_-^2+\alpha_+^2}\Bigr)(S_{1i}^{\ast}S_{1i}+S_{3i}^{\ast}S_{3i})+\Bigl(\frac{\alpha_-^2}{h_-^2+\alpha_-^2}-\frac{\alpha_+^2}{h_-^2+\alpha_+^2}\Bigr)(S_{2i}^{\ast}S_{2i}+S_{4i}^{\ast}S_{4i})\nonumber\\
    &+\Bigl(\frac{h_-\alpha_-}{h_-^2+\alpha_-^2}-\frac{h_-\alpha_+}{h_-^2+\alpha_+^2}\Bigr)(S_{1i}^{\ast}S_{2i}+S_{2i}^{\ast}S_{1i}+S_{4i}^{\ast}S_{3i}+S_{3i}^{\ast}S_{4i})\Bigr].
    \label{spin_final}
\end{align}

\end{widetext}
\bibliography{S_AF_S}
\end{document}